\documentclass{aa}  

\usepackage{graphicx}
\usepackage{txfonts}
\usepackage{xcolor}
\usepackage[colorlinks=true, linkcolor=blue, filecolor=magenta, citecolor=blue, urlcolor=cyan]{hyperref}
\begin{document}

   \title{Efficiency of black hole formation via collisions\\ in stellar systems}

   \subtitle{Data analysis from simulations and observations}

    \author{M.C. Vergara\inst{1}\thanks{E-mail: Marcelo.C.Vergara@uni-heidelberg.de (MV)}
   \and D.R.G. Schleicher\inst{2}\thanks{E-mail: dschleicher@astro-udec.cl (DS)}
   \and A. Escala\inst{3}\thanks{E-mail: aescala@das.uchile.cl (AE)}
   \and B. Reinoso\inst{4}
   \and F. Flammini Dotti\inst{1}
   \and A. W. H. Kamlah\inst{1,5}
   \and \\ M. Liempi\inst{2}
   \and N. Hoyer\inst{6,5}
   \and N. Neumayer\inst{5}
   \and R. Spurzem\inst{1,7,8}
   }

    \institute{Astronomisches Rechen-Institut, Zentrum für Astronomie, University of Heidelberg,
   Mönchhofstrasse 12-14, 69120, Heidelberg, Germany
   \and Departamento de Astronom\'ia, Facultad Ciencias F\'isicas y Matem\'aticas,
   Universidad de Concepcion, Av. Esteban Iturra s/n Barrio Universitario,
   Casilla 160-C, Concepcion, Chile
   \and Departamento de Astronomía, Universidad de Chile, Casilla 36-D, Santiago, Chile
   \and Universität Heidelberg, Zentrum für Astronomie, Institut für Theoretische Astrophysik,
   Albert-Ueberle-Str. 2, 69120 Heidelberg, Germany
   \and Max-Planck-Institut für Astronomie, Königstuhl 17, 69117 Heidelberg, Germany
   \and Donostia International Physics Center, Paseo Manuel de Lardizabal 4,
   E-20118 Donostia-San Sebastián, Spain
   \and National Astronomical Observatories and Key Laboratory of Computational Astrophysics,
   Chinese Academy of Sciences, 20A Datun Rd., Chaoyang District, Beijing 100012, China
   \and Kavli Institute for Astronomy and Astrophysics, Peking University,
   Yiheyuan Lu 5, Haidian Qu, 100871, Beijing, China
   }

   \date{Received September 15, XXXX; accepted March 16, YYYY}

 
  \abstract
   {This paper explores the theoretical relation between star clusters and black holes within them, focusing on the potential role of nuclear star clusters (\textsc{NSCs}), globular clusters (\textsc{GCs}), and ultra-compact dwarf galaxies (\textsc{UCDs}) as environments that allow for black hole formation via stellar collisions.}
   {This study aims to identify the optimal conditions for stellar collisions across a range of stellar systems,  leading to the formation of very massive stars that subsequently collapse into black holes. We analyze data from numerical simulations and observations of diverse stellar systems, encompassing various initial conditions, initial mass functions, and evolution scenarios.}
   {We computed a critical mass, determined by the interplay of the collision time, system age, and initial properties of the star cluster. The efficiency of black hole formation ($\epsilon_{\mathrm{BH}}$) is defined as the ratio of initial stellar mass divided by the critical mass.}
   {We find that stellar systems with a ratio of initial stellar mass over critical mass above 1 exhibit a high efficiency in terms of of black hole formation, ranging from $30-100\%$. While there is some scatter, potentially attributed to complex system histories and the presence of gas, the results highlight the potential for achieving high efficiencies via a purely collisional channel in black hole formation.}
   {In conclusion, this theoretical exploration elucidates the connection between star clusters and black hole formation. The study underscores the significance of \textsc{UCDs}, \textsc{GCs}, and \textsc{NSCs} as environments conducive to the black hole formation scenario via stellar collisions. The defined black hole formation efficiency ($\epsilon_{\mathrm{BH}}$) is shown to be influenced by the ratio of the initial stellar mass to the critical mass.}


   \maketitle
%
\section{Introduction}
The origin of one of the most compact massive objects in the Universe, namely, supermassive black holes (\textsc{SMBHs}) has not been determined thus far. These objects exhibit masses ranging between $10^6-10^{10} ~ \mathrm{M_\odot}$ \citep{Natarajan2009, Volonteri2010, Kormendy2010, Kormendy2013, McConnell2011, Wenwen2015, King2016, Pacucci2017, Mehrgan2019, Onken2020}. Remarkable findings on \textsc{SMBHs} in the distant universe have been published by different research teams, starting from the first three quasars observed at redshift ($z$) higher than 6 reported by \citet{Fan2003}. Afterward, \citet{Mortlock2011} observed a quasar at $z=7.085$ with a mass of $2\times10^9 ~ \mathrm M_\odot$. Soon after, \citet{Wu2015} discovered one of the most massive SMBHs known to date, weighing in at $1.2\times10^{10} ~ \mathrm{M_\odot}$ at $z=6.3$. A few years later, \citet{Jeram2020} found a quasar at $z=2.89$ with a mass of $2.7\times10^{10} ~ \mathrm{M_\odot}$, which is believed to be the most massive SMBH known thus far. Additionally, \citet{Wang2021} contributed to the exploration by spotting a distant active galactic nucleus (AGN) at $z=7.642$, with a mass of $(1.6\pm 0.4)\times10^9 ~ \mathrm{M_\odot}$. \citet{Banados2016} provided the observation of more than $100$ quasars with redshifts between $5.6 \lesssim z \lesssim 6.7$. Most recently, \citet{Fan2023} presented a review of high redshift quasars that includes more than $300$ quasars at $z>6$. 

Some scenarios depict \textsc{SMBH} formation as being related to the remnants of population III stars \citep{Machida2013, Latif2014, Riaz2022}. These stars usually form in metal-free clouds and accumulate mass onto their cores \citep{Omukai1998, Volonteri2003, Tan2004, Ricarte2018}. Another possible route is the direct collapse model, where massive gas clouds collapse within atomic cooling halos at high redshift \citep{Bromm2003, Volonteri2008, Latif2013, Latif2015, Zwick2023} and runaway collisions and mergers of stars take place within star clusters \citep{Rees1984, Portegies2002, Fujii2013, Katz2015, Sakurai2017,  Reinoso2018, Reinoso2020, Sakurai2019, Vergara2021,  Vergara2023}. In addition, some authors have proposed runaway collisions and mergers with gas accretion  to explain their origins \citep{Boekholt2018, Regan2020, Chon2020, Das2021, Schleicher2022, Schleicher2023, Reinoso2023}. Runaway stellar collisions can lead to the formation of a very massive star (\textsc{VMS}), which eventually results in a supernova and/or forms a stellar mass black hole \citep{Heger2002, Whalen2013, Chen2014}. The interaction of the black hole with stars or other stellar mass black holes, besides their capacity to accrete surrounding material, can lead to the formation of an intermediate-mass black hole (\textsc{IMBH}) \citep{Arca-Sedda2023a, Arca-Sedda2023b, Arca-Sedda2023c}. Stellar systems where this process may potentially occur include \textsc{NSCs} \citep{Georgiev2016, Neumayer2020}, \textsc{GCs} \citep{Lutzgendorf2011, Lutzgendorf2012, Lutzgendorf2013}, and \textsc{UCDs} \citep{seth2014, Ahn2017, Ahn2018, Voggel2018}. Regardless of their different ages, sizes, and masses (i.e., density), all of them offer suitable environments for stellar dynamics to unfold. 

As one of the densest stellar configurations known in the Universe, NSCs usually fall in the mass range of $10^5-10^{7} ~ \mathrm{M_\odot}$ \citep{Boker2002, Walcher2005, Cote2006, Boker2010, Georgiev2009, Georgiev2016, Fritz2016, Ordenes2018, Sanchez2019, Neumayer2020}, with a half-mass radius of a few parsecs to a few tens of parsecs \citep{Georgiev2016}. Typically, \textsc{NSCs} contain multiple stellar populations, since they can host old stars, however, they also present star formation \citep{Walcher2005, Rossa2006, Kacharov2018}. Also, they can coexist with \textsc{SMBHs} \citep{Neumayer2012, Georgiev2016, Neumayer2020, Escala2021, Vergara2023}. There are correlations of the \textsc{SMBH} mass with the mass of the galactic bulge \citep{Magorrian1998, Marconi2003, Haring2004}, the \textsc{SMBH} mass and the velocity dispersion of the surrounding stars \citep{Ferrarese2000, Tremaine2002, Gultekin2009}, as well as between the luminosity of the galaxy bulge and the \textsc{NSC} mass \citep{Ferrarese2006, Rossa2006, Wehner2006, Boker2008, Savorgnan2016}. Investigations of stellar orbits surrounding the \textsc{SMBH} at the center of the Milky Way \citep{Ghez2008, Genzel2010, Gillessen2017}, as well as the observation of the shadow of Sagittarius A* \citep{EHT2022} provide evidence that there is a central massive object \citep{Genzel2010, Schodel2014}. The simultaneous presence of \textsc{SMBHs} and \textsc{NSCs} occur in galaxies with masses around $\sim 10^{10}~\mathrm{\mathrm{M_\odot}}$ \citep{Seth2008}. In less massive galaxies, which have less massive \textsc{SMBHs}, direct observations of their dynamical imprint on the \textsc{NSCs} are a challenging task that requires higher spatial resolution than for more massive galaxies at the same distance. Some studies have proposed the absence of a black hole or have set an upper limit on its mass with $M_{\rm BH}<1500-3000~\mathrm{\mathrm{M_\odot}}$ \citep{Gebhardt2001, Merrit2001}. Other examples of \textsc{NSCs} with upper limits for black holes are NGC 300 and NGC 428 \citep{Neumayer2012}. The gravitational influence of the Galactic center has a strong effect on the surrounding gas and stars, causing them to fall to the Galactic center by gradually losing their angular momentum \citep{Shlosman1990, Escala2006}. At high redshifts, the absence of \textsc{SMBHs} at Galactic centers and the abundance of gas lead to even stronger inflows, resulting in the formation of highly dense regions composed of stars and gas \citep{Prieto2016}. Other proposed accretion mechanisms include clump migration through dynamical friction \citep{Tremaine1975, Escala2007, Elmegreen2008, Arca-Sedda2014, Aharon2015} and gravitational torques during galaxy mergers \citep{Barnes1991, Barnes2002, Mayer2010, Newton2013, Wurster2013, Blumenthal2018, Prieto2021}. 

With a half-mass radius of approximately $\sim 10-50 ~ \mathrm{pc}$ \citep{seth2014, Ahn2018, Faifer2017} and masses of $10^6-10^{8} ~ \mathrm{M_\odot}$ \citep{Dabringhausen2009, Dabringhausen2010}, \textsc{UCDs} stand as the densest stellar systems of the Universe. They are known to be old systems, but determining their age and metallicity can be challenging \citep{Chilingarian2008, Francis2012}. Their origins are still unclear, with some theories proposing them to be massive star clusters formed through the merger of smaller \textsc{GCs} \citep{Kroupa1998, Fellhauer2002a, Fellhauer2002b, Urrutia2019} or remnants of tidally stripped dwarf galaxies \citep{Bassino1994, Bekki2003, Pfeffer2013} Examples of these objects include UCD 330, UCD 320 \citep{Voggel2018}, M60-UCD1 \citep{seth2014}, M59c0 \citep{Ahn2017}, M59-UCD-3 \citep{Ahn2018}, and FUCD3 \citep{Afanasiev2018}.

\textsc{} Stellar systems with a lower central density compared to \textsc{NSCs} are known as GCs, with half-mass radii of a few to several dozen parsecs and masses of $10^4-10^6 ~ \mathrm{M_\odot}$. These clusters are predominantly populated by primordial stars, indicating ages of over 10 billion years \citep{Kravtsov2005}. They are found in the halos of galaxies, typically at distances of tens of thousands of light years from the Galactic center. Most of the \textsc{GCs} in the Milky Way originate from a dual process:\ star formation during the early stages of cosmic evolution and the gradual disruption of dwarf galaxies \citep{White1978, Kruijssen2019}. Some examples of \textsc{GCs} are $\mathrm \omega$ Cen, 47Tuc, M15, or M80 \citep{Harris1996}. These stellar systems (\textsc{NSCs}, \textsc{UCDs,} and \textsc{GCs}) provide an ideal environment for stellar collisions to occur. 

In general, clusters with a total mass below $10^5 ~ \mathrm{M_\odot}$ are not very dense, but can form black holes of a relatively low mass, typically amounting to a few percent of the initial mass \citep{Portegies1999, Fujii2013, Koliopanos2017, Arca-Sedda2023a, Arca-Sedda2023b, Arca-Sedda2023c}. Simulations of more massive clusters ($<10^7 ~ \mathrm{M_\odot}$) are computationally expensive. However, by adopting approximations, they can be explored via Fokker-Planck models at higher cluster masses \citep{Lee1987, Quinlan1990, Stone2017}. \citet{Escala2021} proposed that the formation of massive objects in dense stellar systems is regulated by the relation between the collision time in a given system with the age and the relaxation time of that system, as it determines both whether collisions may efficiently occur and whether the system as a whole will be able to readjust to possible changes as a result of collisions. \citet{Vergara2023} performed N-body simulations to probe this scenario further. From the comparison of the collision time with the age of the system, they derived a critical mass, $M_{\mathrm{crit}}$, which is the mass at which the collision timescale becomes equal to the age of the system. This implies that the large majority of stars in the system could then be expected to go through a collision, implying a drastic growth of the most massive objects. Indeed, \citet{Vergara2023} found that the mass of the most massive object systematically increases with the ratio of the initial mass divided by the critical mass ($M_{\mathrm{i}}/M_{\mathrm{crit}}$). In addition, a very similar relation has been found in a preliminary comparison with observed \textsc{NSCs} \citep{Escala2021}.

In this paper, we aim for a more systematic investigation of the same relation reported in \citet{Vergara2023}. We present and compare a range of simulation results from different investigations in terms of $M_{\mathrm{i}}/M_{\mathrm{crit}}$ values and the growth of the most massive object. In particular, the set of simulations analyzed here includes different radial profiles and initial mass functions, as well as models for stellar evolution; for instance, employing realistic simulations of \textsc{GCs}, such as \textsc{dragon II} \citep{Arca-Sedda2023a, Arca-Sedda2023b, Arca-Sedda2023c}. We also  aim to further  refine the comparison with observational data, including additional \textsc{NSCs}, improved age estimates, and additional stellar systems such as \textsc{GCs} and \textsc{UCDs}.

The paper is organised as follows. Section 2 summarizes the methodology and data employed. Section 3 shows the results of our analysis. Section 4 provides a summary of  conclusions drawn from our investigation.

\section{Methodology} \label{section_meth}

Here, we employed and tested the concept of the critical mass ($M_{\mathrm{crit}}$) introduced by \citet{Vergara2023}, which is described in Section~\ref{critmass}. We accounted for the potential presence of an external potential, namely, one that is\ due to the presence of gas in the stellar system. To test this concept, we gathered data from multiple sources of N-body simulations, from which we determined both $M_{\mathrm{crit}}$ and the black hole formation efficiency ($\epsilon_{\mathrm{BH}}$), which is presented in Section~\ref{eff_bh}. These were computed using various initial conditions such as the initial stellar system mass, $M_{\mathrm{i}}$, number of stars, $N$, the half-mass radius, $R_{\mathrm{h}}$, the stellar mass, $m_*$, and radius, $r_*$, besides the mass of the most massive object, $M_{\mathrm{BH}}$, and age, $\tau$, at the end of the simulations. A summary of the parameter space explored by different authors is provided in our Table~\ref{ic_sims} for simulations without any external potential and in Table~\ref{ic_sims_ext} for simulations with external potential. We further employed observations of \textsc{NSCs}, \textsc{GCs}, and \textsc{UCDs} (Table~\ref{ic_obs_NSC}, ~\ref{ic_obs_GC}, and ~\ref{ic_obs_UCD}, respectively). From these data, we use the current age, $\tau$; mass, $M_{\mathrm{i}}$; and effective radius, $R_{\mathrm{eff}}$, of the stellar system, besides the mass of the black hole, $M_{\mathrm{BH}}$. 

\subsection{Critical mass}\label{critmass}

The critical mass, introduced in \citet{Vergara2023} is:
\begin{equation} \label{eq3_mass_crit}
    M_{\mathrm{crit}}=R_{\mathrm{h}}^{7/3}\left(\frac{4\,\pi \,m_*}{3\,\Sigma_0\,\tau \,G^{1/2}}\right)^{2/3},
\end{equation} where $R_{\mathrm{h}}$ is the half-mass radius of the stellar system, $m_*$ is the average mass of a star within the cluster, $\tau$ is the age of the system, $G$ is the gravitational constant, $\Sigma_{\mathrm{0}} = 16 \sqrt{\pi} r_*^2(1+ \Theta)$ is the cross-section for collisions (including gravitational focusing), and $\Theta=9.54 (m_* R_\odot/\rm r_* M_\odot)(~ 100\, kms^{-1}/\sigma)^2$ is the Safronov number (where $r_*$ is the stellar radius and $\sigma$ the velocity dispersion of the stellar system). We evaluated the Safronov number assuming virial equilibrium ($\sigma=\sqrt{GM/R}$), which is commonly employed in the initial conditions of numerical simulations.

In the presence of an external potential due to the presence of gas with a mass, $M_{\mathrm{g}}$, within the stellar system, it is straightforward to generalize this expression under the assumption of virialization with the formalism employed by \citet{Reinoso2020}, namely,

\begin{equation} \label{eq3_mass_crit_ext}
    M_{\rm crit,ext}=R_{\mathrm{h}}^{7/3}\left(\frac{4\,\pi \,m_*}{3\,\Sigma_0\,\tau \,G^{1/2}}\right)^{2/3}\left( 1+q \right)^{-2/3},
\end{equation} with $q=M_{\mathrm{g}}/M_{\mathrm{i}}$. The presence of the external potential thus effectively reduces the critical mass, as it increases the velocity dispersion in the stellar system and therefore increases the likelihood for collisions to occur during the age of the system. 

\subsection{Black hole formation efficiency}\label{eff_bh}

The other central quantity in this paper is the black hole formation efficiency ($\epsilon_{\mathrm{BH}}$), which is defined as the mass of the most massive object formed via collisions divided by the final stellar mass of the stellar system:

\begin{equation}
\epsilon_{\mathrm{BH}}=(1+M_{\rm f}/M_{\rm BH})^{-1}.\label{effbh}
\end{equation} We note, in particular, that the gas mass is not being considered in the definition of $\epsilon_{\mathrm{BH}}$, especially as we are not considering the effects of accretion, whereas the possible effects of the gas are only included through its effect on the steepening of the gravitational potential. A more detailed treatment of the gas would lead to accretion and would make the system more dissipative overall, thereby increasing the masses of the most massive objects to form. However, such a treatment is beyond the scope of the present paper. 

\subsection{Numeric models}\label{num_mo}

In this subsection, we compile the data from different sources of N-body simulations. These simulations were performed using different codes, including \textsc{bridge} \citep{Fujii2007}, \textsc{starlab} \citep{McMillan1996}, \textsc{nbody6} \citep{Aarseth1999b, Aarseth1999}, \textsc{nbody6++gpu} \citep{Wang2015}, and \textsc{bifrost} \citep{Rantala2023}.  These codes have included different models for the star distribution, including profiles corresponding to \citet{Plummer1911, King1966, Miyamoto1975}. They also employ different assumptions concerning the stellar mass function, including models with equal mass stars or employing an initial mass function (IMF) \citep{Salpeter1955, Scalo1986, Kroupa2001}. In the following, we briefly describe the different simulations considered in this investigation.

\citet{Fujii2013} explored the role of stellar collisions in virialized isolated cluster models, using a \citet{Salpeter1955} IMF and the \citet{King1966} density profile. While \citet{Sakurai2017} explored runaway collisions in first star clusters, using the IMF from \citet{Salpeter1955}. The distribution of stars was obtained from cosmological hydrodynamical simulations. Both studies use the \textsc{bridge} code \citep{Fujii2007}

\citet{Portegies1999} investigated runaway collisions in young compact star clusters, utilizing a \citet{Scalo1986} IMF, while \citet{Mapelli2016} explored the impact of metallicity on runaway collisions, employing a \citet{Kroupa2001} IMF. Both investigations employed a \citet{King1966} profile and the {\sc starlab} software \citep{McMillan1996}.

\citet{Katz2015} investigated collisions in primordial star clusters at high redshift, using a \citet{Salpeter1955} IMF. \citet{Reinoso2018} studied collisions in primordial star clusters, employing equal-mass stars. \citet{Reinoso2020} explored runaway collisions in dense star clusters, considering the effects of an external potential. \citet{Katz2015, Reinoso2018, Reinoso2020} used Plummer models \citep{Plummer1911}. \citet{Vergara2021} investigated collisions in flat and rotating clusters using a Miyamoto-Nagai distribution \citep{Miyamoto1975}. All studies used the \textsc{nbody6} code \citep{Aarseth2000}.

\citet{Panamarev2019} simulated the Galactic center of the Milky Way using one million particles, despite the galactic center having approximately more stars by two
orders of magnitude, the computational limitations prevent the inclusion of a higher number of stars, the study employed the IMF from \citet{Kroupa2001}. \citet{Vergara2023} explored runaway collisions in \textsc{NSCs}, using models with equal-mass stars. \citet{Arca-Sedda2023a, Arca-Sedda2023b, Arca-Sedda2023c}, investigated star cluster evolution with up to one million stars, these are some of the most realistic simulations (\textsc{dragon-II}), including a \citet{Kroupa2001} IMF, primordial binaries, stellar evolution, and dynamics of compact objects. All studies use a Plummer distribution \citep{Plummer1911}. The simulations employed the \textsc{nbody6++gpu} code \citep{Wang2015}.
\citet{Rizzuto2023} simulated dense clusters of low mass stars, using a \citet{Kroupa2001} IMF and including an initial central black hole. This work was made with the  \textsc{bifrost} code \citep{Rantala2023}.

The key parameters for computing the critical mass and black hole formation efficiency are summarized in Table~\ref{ic_sims}. In addition, in  Table~\ref{ic_sims_ext}, we show the simulations of \citet{Reinoso2020}, where we include $q=M_{\mathrm{g}}/M_{\mathrm{i}}$. The complete Tables~\ref{ic_sims} and~\ref{ic_sims_ext} are available online (see Appendix~\ref{app_A} for more details).

The \textsc{dragon} simulations \citep{Wang2016} are excluded from this analysis. Although these simulations give rise to multiple black holes, none of the simulations exhibit a dominant mass and therefore do not become the most massive object in the cluster.

All the previously mentioned simulations employ N-body codes to study the stellar dynamics. While they share some similarities, they also have differences. The principal algorithm shared between codes is the Hermite integrator scheme, while \textsc{bridge} uses a sixth-order scheme. The other three codes use a fourth-order scheme. Moreover, \textsc{nbody6} and \textsc{nbody6++gpu} are direct N-body codes and have several algorithms in common, such as the Kustaanheimo-Stiefel regularization to solve close encounters \citep{Stiefel1965}, chain regularization \citep{Mikkola1990, Mikkola1993}, and the Ahmad-Cohen neighbor scheme which spatially splits the hierarchy of the stars to speed up computational calculations \citep{Ahmad1973}. Indeed, \textsc{nbody6++gpu} is the improved version of \textsc{nbody6}, thus \textsc{nbody6++gpu} is optimized with GPU-accelerated supercomputing, which speeds up the calculations. This code also incorporates algorithms for solving stellar dynamics and recipes for stellar evolution proposed by \citet{Hurley2000, Hurley2002}, implemented by \citet{Kamlah2022}. On the other hand, the most recent version of \textsc{starlab} is also a direct N-body code, adopted the stellar evolution recipe by \citet{Hurley2000}; however, in the past, the code used recipes from \citet{Portegies1996} for stellar evolution, incorporating the \citet{Eggleton1989} mass-loss equations and the \citet{Schaerer1999} model for massive stars losing their hydrogen envelope. \textsc{bridge} is a tree-direct hybrid N-body code. The internal region of the cluster is accurately solved through direct integration using the   Hermite scheme \citep{Nitadori2008}, while the external motion is computed employing the tree algorithm \citep{Barnes1986, Makino2004}. The separation between the tree and direct calculations is achieved through Hamiltonian splitting \citep{Kinoshita1990, Wisdom1991}. Finally,  \textsc{bifrost} is a direct summation N-body code accelerated by GPU with a hierarchical formulation of fourth-order forward symplectic integrator, originating in \textsc{frost} \citep{Rantala2021}, the precursor of \textsc{bifrost,} where the integrator was initially derived and tested. We note that the code \textsc{petar} \citep{Wang2020} has been used to study mergers in star clusters \citep{Wang2024, Barber2024} and population III star clusters \citep{Wang2022, Liu2023}. Although we did not include this data, \citet{Wang2022} simulations fall within $M_i/M_{crit} < 0.001$, keeping our conclusions unchanged. More information about algorithms of stellar dynamics is available in the review of computational methods for collisional stellar systems made by \citet{Spurzem2023}.

\begin{table}
    \centering
    \small
    \caption{Initial conditions for simulation sets}
    \begin{tabular}{ccccccc}\hline\hline
       $M_{\mathrm{i}}$& $R_{\mathrm{h}}$& $N$& $M_{\mathrm{BH}}$ & $\tau$ & Code &Ref.\\
       $[{\mathrm{M_\odot}}]$& $[{\rm pc}]$& & $[{\mathrm{M_\odot}}]$ & $[{\rm Myr}]$ & &
       \\\hline
    8233 & 0.25 & 12288 & 140 & 5.5 &SL&P+99\\
    8233 & 0.25 & 12288 & 100 & 4.5&\\
    ... & ... & ... & ... & ...&\\\hline
    6300 & 0.10 & 2048 & 182 & 5&B&F+13\\
    25000 & 0.22 & 8192 & 399 & 5&\\
    ... & ... & ... & ... & ...&\\\hline
    10100 & 0.06 & 14429 & 694.5 & 3.5&NB6&K+15\\
    10100 & 0.11 & 14429 & 494 & 3.5&\\
    ... & ... & ... & ... & ...&\\\hline
    65000 & 0.98 & 100000 & 22 & 17&SL&M+16\\
    65000 & 0.98 & 100000 & 212 & 17&\\
    ... & ... & ... & ... & ...&\\\hline
    164000 & 1.07 & 19900 & 929 & 3&B&S+17\\
    130000 & 0.84 & 15700 & 409 & 3&\\
    ... & ... & ... & ... & ...&\\\hline
    10000 & 0.10 & 100 & 300 & 2&NB6&R+18\\
    10000 & 0.10 & 100 & 300 & 2&\\
    ... & ... & ... & ... & ...&\\\hline
    618000 & 4.20 & 1000000 & 10000 & 5500&NB6+&P+19\\\hline
    10000 & 0.11 & 100 & 350 & 15.6&NB6&R+20\\
    10000 & 0.11 & 100 & 466 & 15.6&\\
    ... & ... & ... & ... & ...&\\\hline
    100000 & 0.08 & 1000 & 6310 & 2&NB6&V+21\\
    100000 & 0.08 & 10000 & 39800 & 2 &\\
    ... & ... & ... & ... & ...&\\\hline
    50000 & 0.008 & 1000 & 9917 & 10&NB6+&V+23\\
    25000 & 0.008 & 500 & 3850 & 10&\\
    ... & ... & ... & ... & ...&\\\hline
    70000 & 1.75 & 120000 & 64 & 2379 &NB6+&AS+23\\
    180000 & 1.75 & 300000 & 69 & 1196&\\
    ... & ... & ... & ... & ...&\\\hline
    89590 & 0.4 & 256000 & 1152 & 41.2&BF&R+23\\
    89590 & 0.6 & 256000 & 1329 & 148\\
    ... & ... & ... & ... & ...&\\\hline
    \end{tabular} \label{ic_sims}
    \tablefoot{Summary of the initial conditions for the different sets of simulations. The first and second columns are Mi and Rh which are the initial mass and half-mass radius of the stellar system, respectively. The third column is the number of stars. The fourth column MBH is the black hole mass. The fifth is the time of the simulation. The sixth column shows the codes (B: bridge, SL: starlab, NB6: nbody6, NB6+: nbody6++gpu, and BF: bifrost) and the last column shows the references. The full table is available at the CDS.} \tablebib{P+99: \citet{Portegies1999}, F+13: \citet{Fujii2013}, K+15: \citet{Katz2015}, M+16: \citet{Mapelli2016}, S+17: \citet{Sakurai2017}, R+18: \citet{Reinoso2018}, P+19: \citet{Panamarev2019}, R+20:  \citet{Reinoso2020}, V+21: \citet{Vergara2021}, V+23: \citet{Vergara2023}, AS+23: \citet{Arca-Sedda2023a, Arca-Sedda2023b, Arca-Sedda2023c}, R+23: \citet{Rizzuto2023}.}
\end{table}   

\begin{table}
    \centering
    \small
    \caption{Initial conditions for simulation sets with external potential}
    \begin{tabular}{cccccccc}\hline\hline
       $M_{\mathrm{i}}$&$R_{\mathrm{h}}$& $N$& $M_{\mathrm{BH}}$ & $\tau$ & q&Code&Ref.\\
       $[{\mathrm{M_\odot}}]$ &$[{\rm pc}]$& & $[{\mathrm{M_\odot}}]$ & $[{\rm Myr}]$& &  &\\\hline
       10000& 0.11 & 100 & 683 & 16.5 &  1.0&NB6&R+20*\\
       10000& 0.11 & 100 & 983 & 16.5 &  1.0&\\
       ... & ... & ... & ... & ...&...\\\hline
    \end{tabular} \label{ic_sims_ext}
    \tablefoot{Summary of the initial conditions for the different sets of simulations with external potential. Same columns as in Table~\ref{ic_sims}, adding an extra column $q$. The full table is available at the CDS.}
    \tablebib{R+20*: \citet{Reinoso2020}}
\end{table} 

\subsection{Observations}

In this subsection, we summarize the observational data on different types of stellar systems we employ in this analysis, considering \textsc{NSCs}, \textsc{UCDs}, and \textsc{GCs}. We summarize the critical parameters, including the age of the cluster, the effective radii, the masses of the stars, and the black hole within in Table~\ref{ic_obs_NSC} (\textsc{NSCs}), Table~\ref{ic_obs_GC} (\textsc{GCs}), and Table~\ref{ic_obs_UCD} (\textsc{UCDs}). We note that for all stellar systems, we assumed that the initial mass is the sum of the current stellar mass and the mass of the black hole\footnote{In principle, additional mass loss could have occured, as we also find in our simulations. We checked that very similar results are obtained if we apply the same correction factor (accounting for mass loss) as we did in the case of simulations. However, in real physical systems, the situation may be more complex; particularly in the case of NSCs, they could also gain mass both due to in situ star formation and mergers with GCs. We therefore prefer to stick with a minimal set of assumptions within the analysis.}.

\begin{table*}
    \centering
    \small
    \caption{Observational properties of \textsc{NSCs}.} 
    \begin{tabular}{llclcrcrc}\hline\hline
    ID &  $M_{\rm BH}~\rm[{\mathrm{M_\odot}}]$& Ref. & $M_{\rm \textsc{NSC}}~\rm[{\mathrm{M_\odot}}]$& Ref. & $R_{\mathrm{eff}}~\rm[{pc}]$& Ref.& $\tau ~\rm[{Myr}]$ & Ref. \\ \hline
    MW & $4.04 \times 10^{6}$ & (1, 2) & $2.50 \times 10^{7}$ & (3) & $4.20$ & (3) & $5.5\times 10^{3}$& (4, 5)\\
    IC342  & $3.16 \times 10^{5}$ &(6)& $1.25 \times 10^{7}$ &(6)& $1.39$ &(6)&$3.4 \times 10^{1}$ &(6)\\
    NGC205 & $6.80 \times 10^{3}$ &(8)& $2.50\ \times 10^{6}$ &(7)& $1.30$ &(7)&$3.0\times10^3$ &(9, 10) \\
    NGC221 & $2.50 \times 10^{6}$ &(7)& $1.70 \times 10^{7}$ &(7)& $4.40$ &(7)&$2.9\times10^3$ &(11) \\
    NGC404 & $5.50\times 10^{5}$ & (12) & $1.10 \times 10^{7}$ & (13) & $15.00$ & (13) & $1.0\times 10^{3}$& (13) \\
    NGC428 & $3.00\times10^4$ &(14)& $2.60\times10^6$ &(14)& $1.20$ &(15)&$1.8\times10^3$ &(16)\\
    NGC1042 & $2.50\times10^4$ &(14)& $3.20\times10^6$ &(14)& $1.30$ &(15)&$1.1\times10^4$ &(16)\\
    NGC1336 & $4.30 \times 10^{7}$ &(17)& $7.00 \times 10^{8}$ &(18)& $66.50$ &(19)&$8.0\times10^3$ &(18)\\
    NGC1493 & $2.50 \times 10^{5}$ &(14)& $3.50 \times 10^{6}$ &(14)& $3.60$ &(15)&$5.7\times10^3$ &(16)\\
    NGC2139 & $1.50 \times 10^{5}$ &(14)& $2.70 \times 10^{7}$ &(15)& $14.80$ &(15)&$4.1\times10^1$ &(16)\\
    NGC2787 & $4.10 \times 10^{7}$ &(20)& $7.00 \times 10^{7}$ &(21)& $5.10$ &(22)&$8.0\times10^3$ &(*)\\
    NGC3115 & $9.10 \times 10^{8}$ &(23)& $7.20 \times 10^{6}$ &(24)& $6.60$ &(22)&$5.0\times10^3$ &(25)\\
    NGC3368 & $7.50 \times 10^{6}$ &(26)& $4.80 \times 10^{7}$ &(27)& $10.40$ &(27)&$3.0\times10^3$ &(28)\\
    NGC3423 & $1.50 \times 10^{5}$ &(14)& $1.90 \times 10^{6}$ &(15)& $2.20$ &(15)&$5.6\times10^3$ &(16)\\
    NGC3593 & $2.40 \times 10^{6}$ &(29)& $1.70 \times 10^{7}$ &(29)& $5.10$ &(22, 29)&$2.0\times10^3$ &(30)\\
    NGC3621 & $3.00 \times 10^{6}$ &(31)& $6.50 \times 10^{6}$ &(15)& $1.80$ &(15)&$1.6\times10^3$ &(32)\\
    NGC4395 & $3.60 \times 10^{5}$ &(33, 34)& $2.00 \times 10^{6}$ &(15, 33)& $1.50$ &(15)&$3.0\times10^3$ &(33)\\
    NGC4414 & $1.50 \times 10^{6}$ &(35)& $1.20 \times 10^{8}$ &(15)& $26.50$ &(15)&$4.5\times10^3$ &(35)\\
    NGC4486 & $6.60 \times 10^{9}$ &(36)& $2.00 \times 10^{8}$ &(14)& $7.40$ &(37)&$8.0\times10^3$ &(*)\\
    NGC4501 & $1.90 \times 10^{7}$ &(26)& $5.20 \times 10^{8}$ &(27)& $30.00$ &(27)&$7.5\times10^3$ &(38)\\
    NGC4578 & $3.50 \times 10^{7}$ &(39)& $5.30 \times 10^{7}$ &(40)& $7.03$ &(40)&$8.0\times10^3$ &(*)\\
    NGC4623 & $1.60 \times 10^{8}$ &(41)& $1.30 \times 10^{8}$ &(40)& $48.03$ &(40)&$8.0\times10^3$ &(*)\\
    NGC4697 & $1.70 \times 10^{8}$ &(42)& $2.80 \times 10^{7}$ &(24)& $4.40$ &(24)&$8.9\times10^3$ &(43)\\
    NGC4699 & $1.54 \times 10^{10}$ &(26)& $6.30 \times 10^{8}$ &(27)& $25.75$ &(27)&$8.0\times10^3$ &(*)\\
    NGC5055 & $8.50 \times 10^{8}$ &(20)& $1.70 \times 10^{6}$ &(22)& $13.50$ &(22)&$3.8\times10^3$ &(28)\\
    NGC5102 & $9.10 \times 10^{5}$ &(8)& $7.30 \times 10^{7}$ &(7)& $26.30$ &(7)&$6.9\times10^2$ &(32)\\
    NGC5206 & $6.30 \times 10^{5}$ &(8)& $1.50 \times 10^{7}$ &(7)& $8.10$ &(7)&$2.9\times10^3$ &(32)\\
    NGC7424 & $1.50 \times 10^{5}$ &(14)& $1.00 \times 10^{6}$ &(14, 15)& $6.80$ &(14, 15)&$1.3\times10^3$ &(16)\\
    NGC7713 & $7.50 \times 10^{6}$ &(44)& $4.04 \times 10^{5}$ &(22)& $1.14$ &(22)&$8.0\times10^3$ &(*)\\
    VCC1254 & $9.00 \times 10^{6}$ &(45)& $1.10 \times 10^{7}$ &(24)& $49.20$ &(24)&$5.7\times10^3$ &(46)\\ \hline
    \end{tabular}    \label{ic_obs_NSC}
    \tablefoot{The columns are: (1) stellar system ID; (2, 3) \textsc{BHs} mass and reference; (4, 5) \textsc{NSCs} mass and reference; (6, 7) \textsc{NSCs} effective radius and reference; and (8, 9) age and reference. Note:\ for stellar systems where the age is not defined, we assumed a typical age, denoted by (*).}
    \tablebib{(1) \citet{GC2018}, (2) \citet{Do2019}, (3) \citet{Schodel2014}, (4) \citet{Blum1996}, (5) \citet{Pfuhl2011}, (6) \citet{Boker1999}, (7) \citet{Nguyen2018}, (8) \citet{Nguyen2019}, (9) \citet{Cappellari1999}, (10) \citet{Davidge2003}, (11) \citet{Villaume2017}, (12) \citet{Davis2020}, (13) \citet{Nguyen2017}, (14) \citet{Neumayer2012}, (15) \citet{Georgiev2016}, (16) \citet{Walcher2006}, (17) \citet{Thater2023}, (18) \citet{Fahrion2019}, (19) \citet{Turner2012}, (20) \citet{Graham2008}, (21) \citet{Sarzi2001}, (22) \citet{Pechetti2020}, (23) \citet{Emsellem1999}, (24) \citet{Graham2009}, (25) \citet{Shcherbakov2014}, (26) \citet{Saglia2016}, (27) \citet{Ashok2023}, (28) \citet{Sarzi2005}, (29) \citet{Nguyen2022}, (30) \citet{Coccato2013}, (31) \citet{Barth2009}, (32) \citet{Kacharov2018}, (33) \citet{denBrok2015}, (34) \citet{Peterson2005}, (35) \citet{Thater2017}, (36) \citet{Gebhardt2011}, (37) \citet{Gnedin2014} , (38) \citet{Repetto2017}, (39) \citet{Krajnovic2018}, (40) \citet{Cote2006}, (41) \citet{Pechetti2017}, (42) \citet{Gebhardt2003}, (43) \citet{Trager2000}, (44) \citet{Fusco2022}, (45) \citet{Geha2002} (46) \citet{ Paudel2011}.}
\end{table*}

\begin{table*}
    \centering
    \small
    \caption{Observational properties of \textsc{GCs}. }
    \begin{tabular}{llclcrcrc}\hline\hline
    ID &  $M_{\rm BH}~\rm[{\mathrm{M_\odot}}]$& Ref. & $M_{\rm \textsc{GC}}~\rm[{\mathrm{M_\odot}}]$& Ref. & $R_{\mathrm{eff}}~\rm[{pc}]$& Ref.& $\tau ~\rm[{Myr}]$ & Ref. \\ \hline
    G1       & $2.30 \times 10^4$ & (1) & $5.75 \times 10^6$ & (2) & $6.50$ & (3) & $13.0 \times 10^3$ & (13) \\
    NGC104  & $1.50 \times 10^3$ & (4) & $1.10 \times 10^6$ & (1) & $4.10$ & (5) & $10.7 \times 10^3$ & (14) \\
    NGC1851 & $2.00 \times 10^3$ & (6) & $3.72 \times 10^5$ & (6) & $1.80$ & (5) & $8.8 \times 10^3$ & (14) \\
    NGC1904 & $3.00 \times 10^3$ & (6) & $1.41 \times 10^5$ & (6) & $2.40$ & (5) & $ 9.4\times 10^3$ & (14) \\
    NGC2808 & $1.00 \times 10^4$ & (7) & $8.13 \times 10^5$ & (7) & $2.20$ & (5) & $8.2 \times 10^3$ & (14) \\
    NGC5139 & $4.70 \times 10^4$ & (8) & $2.51 \times 10^6$ & (8) & $7.60$ & (5) & $12.0 \times 10^3$ & (*) \\
    NGC5286 & $1.50 \times 10^3$ & (9) & $2.82 \times 10^5$ & (9) & $2.50$ & (5) & $12.0 \times 10^3$ & (*) \\
    NGC5694 & $8.00 \times 10^3$ & (6) & $2.57 \times 10^5$ & (6) & $4.10$ & (5) & $11.4 \times 10^3$ & (14) \\
    NGC5824 & $6.00 \times 10^3$ & (6) & $4.47 \times 10^5$ & (6) & $4.20$ & (5) & $10.9 \times 10^3$ & (14) \\
    NGC6093 & $8.00 \times 10^2$ & (6) & $3.39 \times 10^5$ & (6) & $1.80$ & (5) & $13.0 \times 10^3$ & (15) \\
    NGC6266 & $2.00 \times 10^3$ & (6) & $9.33 \times 10^5$ & (6) & $1.80$ & (5) & $10.0 \times 10^3$ & (14) \\
    NGC6388 & $1.70 \times 10^4$ & (10) & $1.10 \times 10^6$ & (10) & $1.50$ & (5) & $12.0 \times 10^3$ & (*) \\
    NGC6715 & $9.40 \times 10^3$ & (11) & $1.91 \times 10^6$ & (1) & $6.30$ & (5) & $12.0 \times 10^3$ & (*) \\
    NGC7078 & $4.40 \times 10^3$ & (12) & $6.17 \times 10^5$ & (12) & $3.00$ & (5) & $10.3 \times 10^3$ & (14) \\\hline
\end{tabular}    \label{ic_obs_GC}
\tablefoot{The columns are the same as in Table~\ref{ic_obs_NSC}.}
\tablebib{(1) \citet{McLaughlin2006}, (2) \citet{Gebhardt2005}, (3) \citet{Ma2007}, (4) \citet{Lutzgendorf2013b}, (5) \citet{Harris1996}, (6) \citet{Lutzgendorf2013a}, (7) \citet{Lutzgendorf2012}, (8) \citet{Noyola2011}, (9) \citet{Feldmeier2013}, (10) \citet{Lutzgendorf2011}, (11) \citet{Ibata2009}, (12) \citet{Gerssen2002}, (13) \citet{Meylan2001}, (14) \citet{DeAngeli2005}, (15) \citet{Rosenberg1999}.}
\end{table*}

\begin{table*}
    \centering
    \small
    \caption{Observational properties of \textsc{UCDs}.}
    \begin{tabular}{llclcrcrc}\hline\hline
    ID &  $M_{\rm BH}~\rm[{\mathrm{M_\odot}}]$& Ref. & $M_{\rm \textsc{UCD}}~\rm[{\mathrm{M_\odot}}]$& Ref. & $R_{\mathrm{eff}}~\rm[{pc}]$& Ref. & $\tau ~\rm[{Myr}]$ & Ref. \\ \hline
    B023-G078 & $9.10 \times 10^{4}$ &(1)& $6.22 \times 10^{6}$ &(1)& $18.70$ &(1)&$10.5\times10^3$ &(1)\\
    FUCD3 & $3.30 \times 10^{6}$ &(2)& $8.30 \times 10^{7}$ &(2)& $19.40$ &(2)&$13.9\times10^3$ &(10)\\
    M59c0 & $5.80 \times 10^{6}$ &(3)& $8.30 \times 10^{7}$ &(3)& $32.00$ &(3)&$9.3\times10^3$ &(9)\\
    M59-UCD-3 & $4.20 \times 10^{6}$ &(4)& $1.90 \times 10^{8}$ &(4)& $27.00$ &(4)&$7.7\times10^3$ &(7)\\
    M60-UCD1 & $2.10 \times 10^{7}$ &(5)& $1.20 \times 10^{8}$ &(5)& $47.90$ &(5)&$10.0\times10^3$ &(*)\\
    UCD330 & $1.00 \times 10^{5}$ &(6)& $6.10 \times 10^{6}$ &(6)& $3.11$ &(6)&$10.0\times10^3$ &(*)\\
    UCD320 & $1.00 \times 10^{6}$ &(6)& $2.81 \times 10^{6}$ &(6)& $4.67$ &(6)&$10.0\times10^3$ &(*)\\
    VUCD3 & $4.40 \times 10^{6}$ &(3)& $6.60 \times 10^{7}$ &(3)& $18.00$ &(3)&$13.0\times10^3$ &(8)\\
    \hline
\end{tabular}    \label{ic_obs_UCD}
\tablefoot{ The columns are the same as in Table~\ref{ic_obs_NSC}.}
\tablebib{(1) \citet{Pechetti2022}, (2) \citet{Afanasiev2018}, (3) \citet{Ahn2017}, (4) \citet{Ahn2018}, (5) \citet{seth2014}, (6) \citet{Voggel2018}, (7) \citet{Villaume2017}, (8) \citet{Francis2012}, (9) \citet{Chilingarian2008}, (10) \citet{Chilingarian2011}.}
\end{table*}

We assumed that the initial mass of the galactic center was the combined mass of the black hole and \textsc{NSC} \citep{Vergara2023}. Additionally, we considered the initial effective radius to be one-tenth of its current size, as suggested by \citet{Banerjee2017}. We assumed that the stars have solar properties (i.e., $1~\rm{\mathrm{M_\odot}}$ and $1~\rm{R_\odot}$). If the \textsc{NSCs} ages were not available we assume a typical age of $8\,$Gyr.

For the \textsc{GCs,} we made the same assumption for the initial mass as for \textsc{NSC} (i.e., $M_{\rm i}=M_{\rm \textsc{GC}}+M_{\rm BH}$). These stellar configurations are one of the oldest objects in the Universe. The effective radius of isolated \textsc{GCs} remains constant through several relaxation timescales \citep{Spitzer1972}, then we assume that the current size mirrors the initial size. \textsc{GCs} are old stellar systems. If the age was not available, we assumed a typical age of $12\,$Gyr.

We analyzed the \textsc{UCDs} by taking into account assumptions from both \textsc{GCs} and \textsc{NSCs} regarding their effective radius; for the initial mass, we consider $M_{\rm i}=M_{\rm \textsc{UCD}}+M_{\rm BH}$, given the uncertain nature of their formation. Some theories propose that \textsc{UCDs} could be remnants of tidal disruptions in dwarf galaxies interacting with larger ones, resulting in the stripping of outer layers and the formation of \textsc{NSCs} \citep{Bassino1994, Bekki2003, Pfeffer2013}. Alternatively, there is a suggestion that \textsc{UCDs} may originate from the merger of multiple \textsc{GCs} \citep{Kroupa1998, Fellhauer2002a, Fellhauer2002b, Urrutia2019}. In our investigation of \textsc{UCDs}, we considered two scenarios. In particular, UCD1 assumes a fixed radius during evolution, incorporating an external potential with $80\%$ dark matter due to observed high mass-to-light ratio values \citep{Baumgardt2008}; UCD2, aligning more with \citet{Fellhauer2006}, posits an absence of dark matter and envisions an initial expansion like \textsc{NSCs}. 
Determining the ages of \textsc{UCDs} is a challenging task \citep{Francis2012}. For those stellar systems that do not present an age measurement, we assumed a typical age in between \textsc{GCs} and \textsc{NSCs} of $10\,$Gyr.

\section{Data analysis and comparison}

In this section, we use the available data on simulations and observations to systematically test the concept of the critical mass. For this purpose, we first briefly recall the main differences between the respective datasets. In the case of simulations, we recall they represent idealized isolated systems. In most cases in the literature, their initial conditions, including the initial stellar mass, $M_{\mathrm{i}}$, have been reported, as well as the main outcomes, such as the mass $M_{\mathrm{BH}}$ of the massive object that forms. In the case of observational data, on the other hand, we know about the current state of the system, namely, the final stellar mass, $M_{\mathrm{f}}$, and the mass of the most massive object, $M_{\mathrm{BH}}$. We also note that real observational systems may be more complex than the simulation and may include additional physical processes. In the case of \textsc{NSCs}, gas physics, star formation, and mergers with \textsc{GCs} could play some role in their evolution. For \textsc{GCs} and \textsc{UCDs}, on the other hand, it is more plausible that the stellar systems might be evolving for very long timescales essentially in isolation and without active star formation. 

The computation of the critical mass (as defined via Eq.~\ref{eq3_mass_crit_ext}) does, in principle, require knowledge of the initial conditions and is relatively trivial in the case of numerical simulations, while in the case of observations some assumptions or estimates may be necessary to determine the critical mass scale. On the other hand, the efficiency parameter $\epsilon_{\mathrm{BH}}$ defined in Eq.~\ref{effbh} requires knowledge of the final stellar mass and the mass of the most massive object. This appears straightforward from an observational perspective, whereas in simulations, the final stellar mass has not always been documented and may thus require an estimate. In the following subsections, we first analyse the results from simulations and observations separately and subsequently attempt to make a comparison between simulation and observation results.

\subsection{Simulations data}

\begin{figure*}
    \centering   
    \includegraphics[]{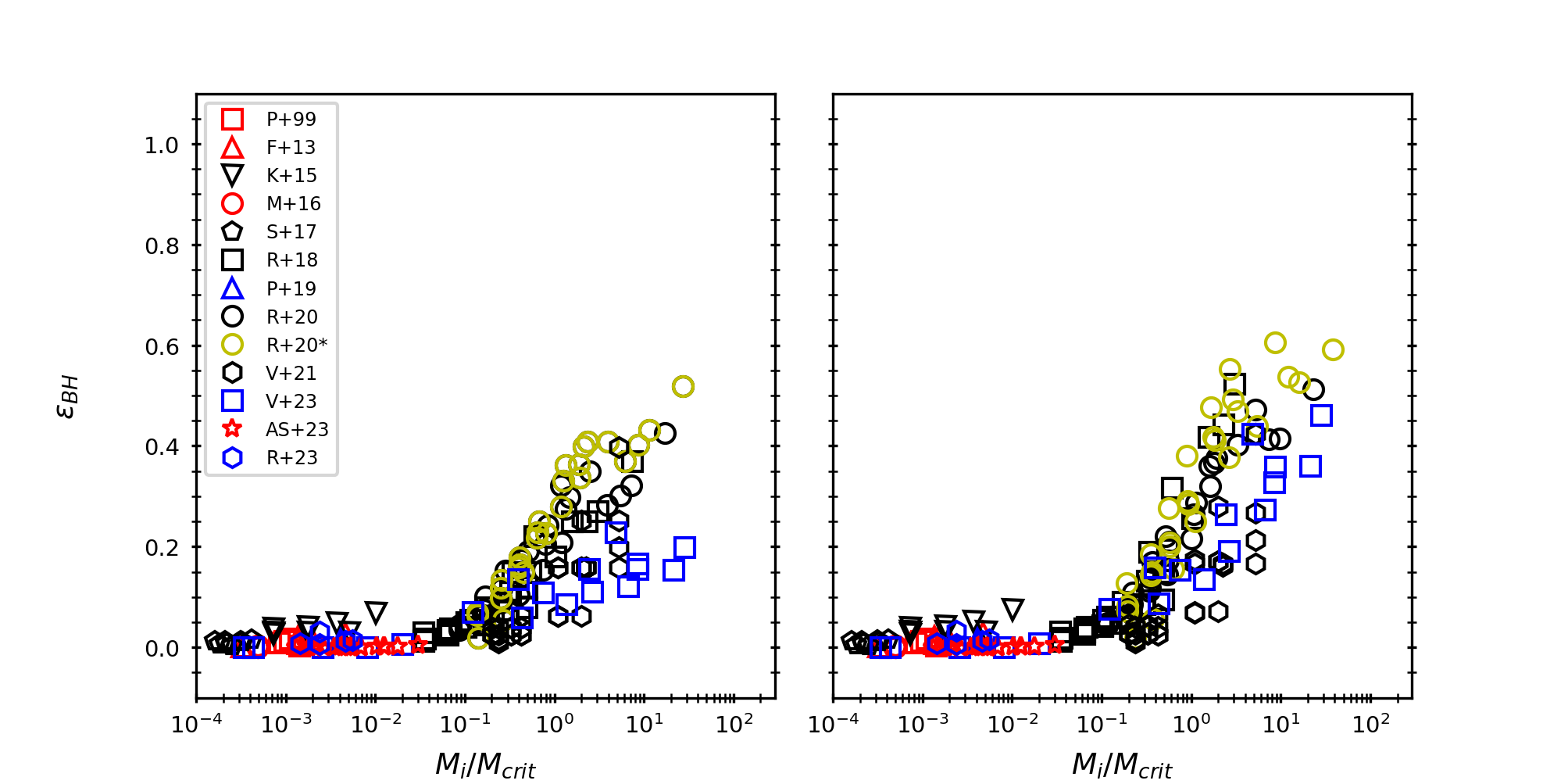}
    \caption{Black hole formation efficiency, $\epsilon_{\mathrm{BH}}$, (left) computed by Eq.~\ref{effbh} against the initial mass of the cluster, $M_{\mathrm{i}}$, normalized by the critical mass, $M_{\mathrm{crit}}$, calculated via Eq.~\ref{eq3_mass_crit}. Different types of star clusters are represented by distinct colors: black symbols for the first star clusters (yellow symbols data from \citet{Reinoso2020}, represent models with external potential, calculated via Eq.~\ref{eq3_mass_crit_ext}.), red symbols for \textsc{GCs} and blue symbols for \textsc{NSCs}. The final stellar mass is estimated neglecting mass loss via Eq.~\ref{noloss}. Right: The black hole formation efficiency $\epsilon_{\mathrm{BH}}$ computed by Eq.~\ref{effbh} against the current mass of the cluster, $M_{\mathrm{i}}$, normalized by the critical mass, $M_{crit}$, calculated from Eq.~\ref{eq3_mass_crit}, including an approximate correction for mass loss.}
    \label{figure1}
\end{figure*}

In simulations, the initial conditions are known, and the calculation of the critical mass, $M_{\mathrm{crit}}$, and the ratio, $M_{\mathrm{i}}/M_{\mathrm{crit}}$, is thus straightforward. Information about the final stellar mass is, in principle, part of the simulations, but  has not usually been reported in the literature. A first simple estimate can be obtained if the mass loss is neglected, but still considering the mass that has gone into the most massive object, implying: \begin{equation}
M_{\mathrm{f}}=M_{\mathrm{i}} - M_{\mathrm{BH}}.\label{noloss}
\end{equation}
The result of this calculation is given in the left panel of Fig.~\ref{figure1}. While we found scatter in the plot, we can  clearly note that higher efficiencies, $\epsilon_{\mathrm{BH}}$, are obtained in all simulations with $M_{\mathrm{i}}/M_{\mathrm{crit}}\gtrsim0.3$ with an increasing trend, the slope of this trend depending somewhat on the set of simulations being considered, with maximum efficiencies reaching around $40-50\%$. On the other hand, for low values of $M_{\mathrm{i}}/M_{\mathrm{crit}}$, the relation appears essentially flat with efficiencies (at best) on the level of a few percent. 

In the left panel of Fig.~\ref{figure1}, mass loss via escapers has been completely neglected, while (in principle) it is clear that it should be considered. We know however at least from some simulations how much mass has been lost during evolution. In Appendix~\ref{ClusterMassLoss}, we employ the simulations of \citet{Vergara2021, Vergara2023, Reinoso2020} to analyze how the ratio $M_{\mathrm{f}}/M_{\mathrm{i}}$ depends on $M_{\mathrm{i}}/M_{\mathrm{crit}}$. Such a dependence can be expected for two reasons: a) if the mass of the most massive object increases with $M_{\mathrm{i}}/M_{\mathrm{crit}}$ as indicated via Fig.~\ref{figure1}, this would imply such a relation even in the absence of escapers; b) escapers occur due to interactions within the stellar system and a higher ratio of $M_{\mathrm{i}}/M_{\mathrm{crit}}$ implies that such interactions should be more frequent. In the appendix, we examine the dependence between these quantities both for the case when escapers are neglected (as when employing Eq.~(\ref{noloss})) and when the full information from the simulations is being employed. We derived a correction factor $\alpha$ (Eq.~\ref{alpha}) to estimate the expected mass loss due to escapers. Of course, we note that the exact volume of mass loss may depend on more factors including the adopted density profile and so the estimate arrived here can at best serve as an approximation. 

Applying this correction factor may then lead to an improved estimate of the final stellar mass, $M_{\mathrm{f}}$, and an improved calculation of $\epsilon_{\mathrm{BH}}$. The result of this calculation is given in the right panel of Fig.~\ref{figure1}. The behavior is qualitatively similar to what is seen in the previous plot, but overall it appears more pronounced. The efficiency parameter $\epsilon_{\mathrm{BH}}$ now reaches up to $60\%$ for $M_{\mathrm{i}}/M_{\mathrm{crit}}\gtrsim1$, although this may also have more moderate values, on the order of $30-40\%$. An increase in the efficiency already appears visible from $M_{\mathrm{i}}/M_{\mathrm{crit}}\gtrsim0.1$. For lower values of that ratio, the results appear very similar as before with efficiencies at best on the percent level. Therefore, the critical mass scale is probably not the only factor regulating the efficiency of collisions, although it still appears to be a very important component regulating this process.

\subsection{Observational data} \label{obs_data}

\begin{figure}
    \centering
    \includegraphics[width=1\linewidth]{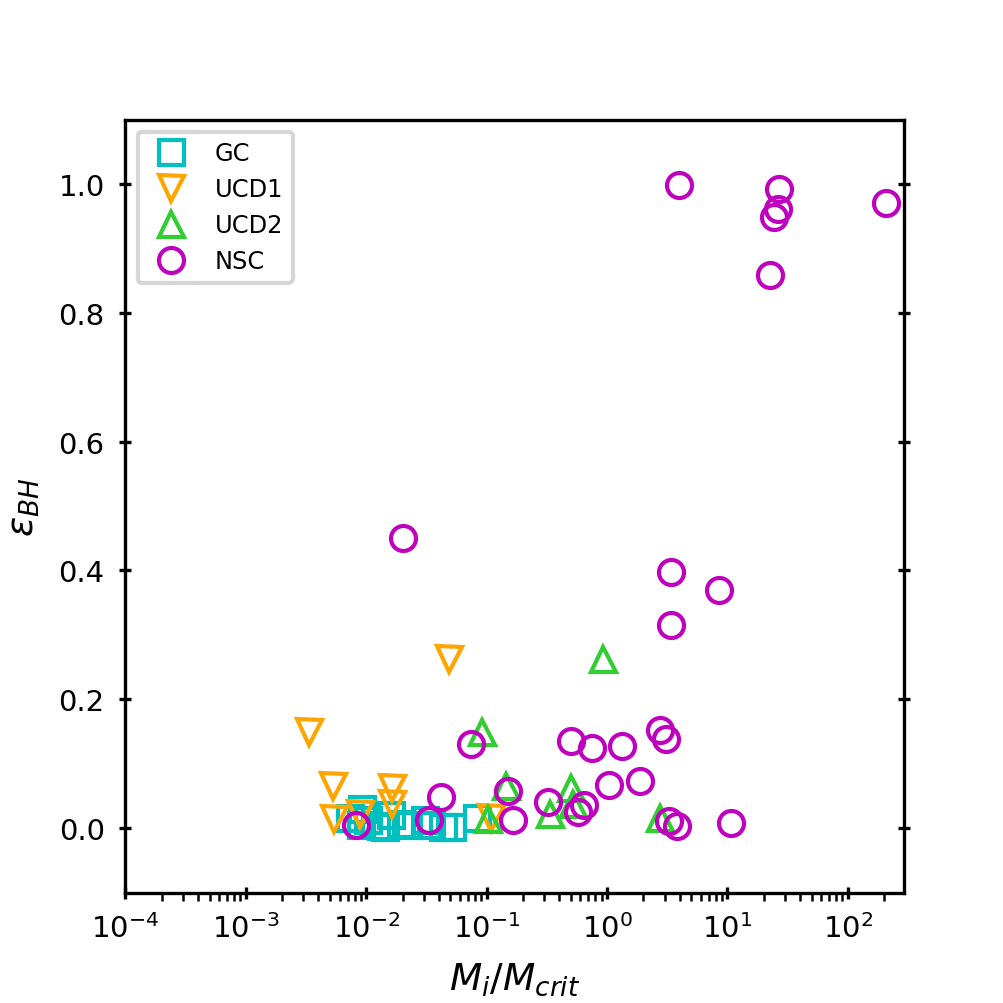}
    \caption{Black hole formation efficiency, $\epsilon_{BH}$, computed via Eq.~\ref{effbh} against the initial mass of the cluster normalized by the critical mass, $M_{crit}$ (estimated from the current properties of the cluster), for observational data.}
    \label{fig:figure2}
\end{figure}

For observations, as already noted above, we input the current conditions of the stellar system including $M_{\mathrm{f}}$ and $M_{\mathrm{BH}}$, allowing for the direct calculation of $\epsilon_{\mathrm{BH}}$. However, it is to be noted that the initial properties for observational data are more uncertain as in the case of simulations, which are meant to model ideal and isolated systems; real stellar clusters may not be isolated and some of them (particularly the \textsc{NSCs}) may even experience recent or ongoing stellar formation and/or mergers with \textsc{GCs}. It is thus a relevant caveat that needs to be taken  (at least for the \textsc{NSCs),} namely, that the evolutionary history, including mass evolution, will be more complex as compared to the simulations. As discussed already in Section~\ref{num_mo}, for \textsc{NSCs,} we expect and assume an expansion during the evolution, while \textsc{GCs} are typically assumed to have approximately constant radii. However, we note that the half-mass radius could increase by a factor of $2-3$ during the evolution \citep{Gieles2010}. The situation of \textsc{UCDs} is less clear and thus, we consider two possible scenarios below. To estimate the initial mass for observations, we can consider it as the sum of the current stellar mass and black hole mass it contains, under the assumption that at least its order of magnitude will be similar to $M_{\mathrm{i}}/M_{\mathrm{crit}}$ for simulations. The results are shown is given in Fig.~\ref{fig:figure2}. The scatter in this figure is somewhat increased, which given the previous considerations is perhaps  unsurprising. Nonetheless, we note that for $M_{\mathrm{i}}/M_{\mathrm{crit}}\gtrsim0.03$, $\epsilon_{\mathrm{BH}}$ is covering the entire range of efficiencies from the percent level up to $\sim60\%$. On the other hand, for low values of $M_{\mathrm{i}}/M_{\mathrm{crit}}$, we only have efficiencies of up to $20\%$ at most. In the case of \textsc{UCDs,} we explore one scenario (UCD1), where they are assumed to have fixed radius during their evolution and considering an external potential with  $80\%$ of dark matter due to the observed mass to high ratios \citep[e.g.,][]{Baumgardt2008}. We also explore another case (UCD2), more compatible with \citet{Fellhauer2006}, in the absence of dark matter. There, we assume that an initial expansion of the systems takes place, as in the case of \textsc{NSCs}. We find that  the  case of UCD2 in particular fits very well into the picture and displays a quite similar behavior as the \textsc{NSCs}. The assumptions made in the case of UCD1 would somewhat enhance the scatter, whereas it does not fundamentally change the picture. We also note that high efficiencies close to $100\%$ are only found when the ratio of stellar mass over critical mass is above $1$; otherwise, it typically ranges from low efficiencies up to the order of $20\%$. We further note that we do not know the physics of how the massive black holes in these systems have formed, as it is at least conceivable that also gas physics should have been involved. The latter may contribute to adding scatter in the plot and increase the efficiencies above the expectation from a purely collisional channel. In particular, when considering the group of \textsc{GCs}, we note that they all behave very similarly; in terms of the physics, it is perhaps also the case that is most clear, as they are expected to evolve essentially in isolation after their initial formation. On the other hand, the \textsc{NSCs} are very likely the most complex systems we consider here, with possibly different formation histories, their contribution very considerably increases the scatter we find in this plot. Thus, the \textsc{UCDs} might be somewhere in between \textsc{GCs} and \textsc{NSCs} judging from their respective scatter within the plot. 

To assess the sensitivity of this plot to the aforementioned assumptions, we estimated $M_{\mathrm{i}}$ as the sum of the current stellar mass and mass of the black hole within. The critical mass range varies depending on the stellar system age and their initial properties. For NSCs, it typically falls within $10^5-10^9 ~\rm{\mathrm{M_\odot}}$, for GCs, it is between $10^6-10^8 ~\rm{\mathrm{M_\odot}}$, and for UCDs, it ranges from $10^9-10^{11} ~\rm{\mathrm{M_\odot}}$ (UCD1) or $10^7-10^9 ~\rm{\mathrm{M_\odot}}$ (UCD2). The results given in Fig.~\ref{fig:figure2} are in particular overall very similar to the left panel of Fig.~\ref{figure1}. We note that for low values of $M_{\mathrm{i}}/M_{\mathrm{crit}}$, efficiencies can reach at best up to $20\%$, while for $M_{\mathrm{i}}/M_{\mathrm{crit}}\gtrsim0.03$ the whole range of efficiencies up to $100\%$ is possible. Our qualitative conclusions are thus similar though the evolutionary history is more complex as compared to the simulations.

We note three special cases: VCC1254, which introduces a higher scatter at $M_{\mathrm{i}}/M_{crit}\sim 10^{-2}$, with $\epsilon_{\mathrm{BH}}$ values around $50\%$. However, the black hole mass of VCC1254 is an upper limit. Besides having a larger nuclear mass-to-light ratio, which means that it must contain an older stellar population \citep{Geha2002}, in contrast to observations by \citet{Durrell1997} and \citet{Stiavelli2001}, which show the presence of young stars. Thus more precise measurements will lead to a more accurate determination of the mass and thus lower $\epsilon_{\mathrm{BH}}$. The \textsc{UCDs} M60-UCD1 and UCD320 show $\epsilon_{\mathrm{BH}}\approx 15-30\%$ at $M_{\mathrm{i}}/M_{crit}\sim 10^{-3}-10^{-2}$, respectively, under the assumptions of UCD1. Compared to the stellar mass, its high black hole mass suggests that its origin could be a tidal core of a dwarf galaxy \citep{Bassino1994, Bekki2003, Pfeffer2013, seth2014, Afanasiev2018}, which aligns with the case of UCD2. Finally, the scatter presented in some observations must be related to their complex formation history, which is beyond the scope of this investigation.

Finally we also note the case of $\mathrm \omega$ Cen (NGC 5139). This \textsc{GC} has been widely studied due to its unique characteristic of being the most massive \textsc{GC} of the Milky Way. Several studies have suggested that the stellar dynamics in $\mathrm \omega$ Cen could be explained due to the presence of an \textsc{IMBH} with a mass $\sim 4\times 10^4~\rm{\mathrm{M_\odot}}$ at the center \citep{Noyola2008, Noyola2011, Jalali2012, Baumgardt2017}. Others have argued that the stellar kinematics can be explained by the presence of a dense cluster of stellar-mass black holes \citep{Baumgardt2019, Zocchi2019}. Other studies suggest that the presence of an \textsc{IMBH} is not necessary to explain the stellar dynamics \citep{Marel2010}. In this investigation, we consider the first option to determine an upper limit for the efficiency. We note that for the other two options, the black hole formation efficiency will be $\epsilon_{\mathrm{BH}}\simeq 0$, as this result still aligns with the trend.

\subsection{Comparative analysis of observational data and simulations data}

So far, we have independently analyzed the data from simulations and observations to check whether they support the concept of a critical mass that significantly determines the efficiency of collisions in dense stellar systems. We have noted the main differences regarding the typical data that are reported for the respective simulations and observations and developed the tools to at least approximately estimate the quantities that cannot be directly inferred. During the analysis of the observational data, we further encountered some of the systematic differences between simulations that usually model idealized and isolated stellar systems. However, real stellar systems, in particular \textsc{NSCs}, are likely more complex due to their star formation histories as well as occasional mergers with \textsc{GCs}. On the other hand, the \textsc{GCs} appear to at least more closely resemble the behavior of an isolated system as their star formation terminated very long ago, while \textsc{UCDs} are somewhat more uncertain as their origin is not known. 

With these differences in mind, we aim to approximately assess what simulations and observations may have in common and which differences can be identified from the comparison. This may, on one hand, allow us to establish processes that play an important role in both observations and simulations, but also help to identify processes that are not yet included in simulations, which may play a possibly important role during their dynamical evolution. 

\begin{figure}
    \centering
    \includegraphics[width=.99\linewidth]{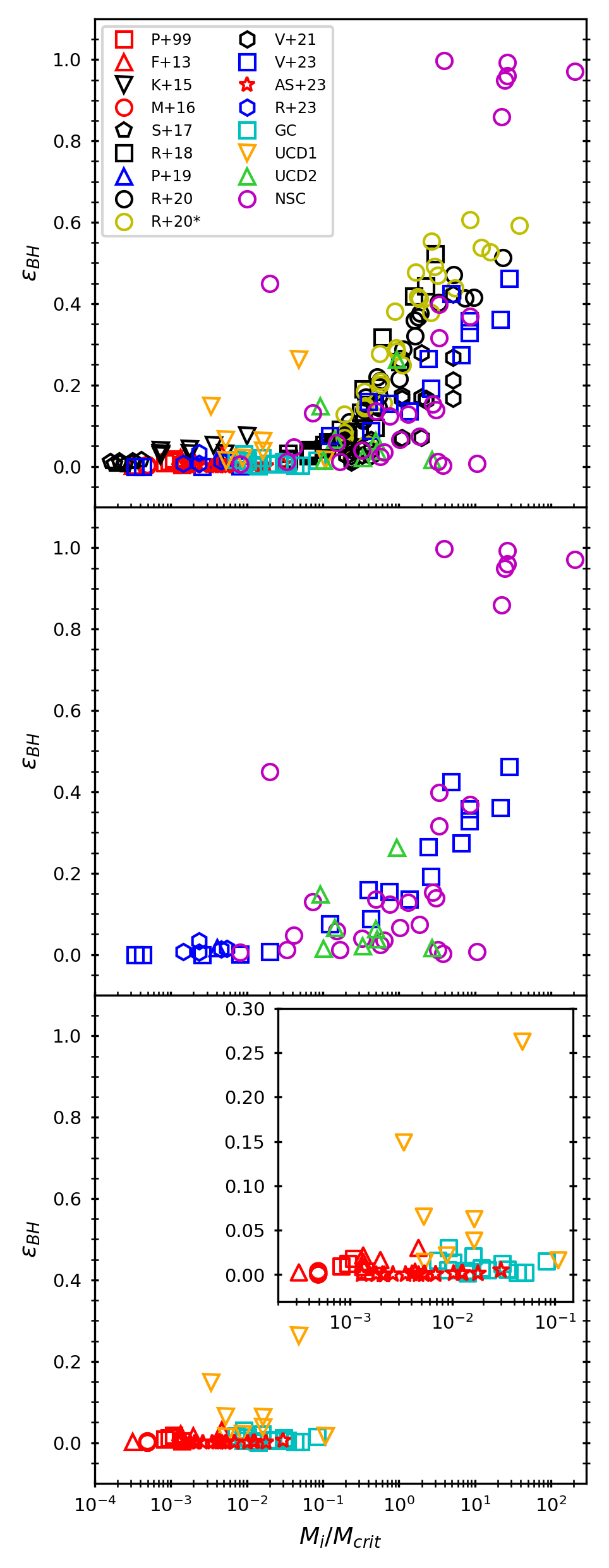}
    \caption{Black hole formation efficiency $\epsilon_{\mathrm{BH}}$ computed by Eq.~\ref{effbh} against the current mass of the cluster $M_{\mathrm{i}}$ normalized by the critical mass $M_{crit}$ (Eq.~\ref{eq3_mass_crit}) for all observations and simulations (top); for observations and simulations of NSC, also including observations of UCD2 (middle); and for observations and simulations of GC, also including observations of UCD1 (bottom).}
    \label{fig:figure3}
\end{figure}

For this analysis here, in the context of the simulations, the mass loss will be estimated using a correction factor, $\alpha,$ (see Eq.~\ref{alpha}) when calculating $\epsilon_{\mathrm{BH}}$. In the observations, we have all the information on the current stage. The result of this calculation is given in Fig.~\ref{fig:figure3}, where the top panel shows the data comparing all simulations and observations, while in the mid and bottom panels, we focus on data (simulations and observations) for \textsc{NSCs} and  \textsc{GCs}, respectively. We note that in the mid panel, we add observations of \textsc{UCDs}, under the assumption of no dark matter but expansion (UCD2), while in the bottom panel, we add observations of \textsc{UCDs}, under the assumption of the presence of dark matter and no expansion (UCD1).

In general, there is a broad consistency between the data from simulations and observations. Where the efficiencies are on the percent level when we consider $M_{\mathrm{i}}/M_{\mathrm{crit}}$ to be low, and only for $M_{\mathrm{i}}/M_{\mathrm{crit}}\gtrsim0.1$ we find higher efficiencies up to of the order $100\%$, but with significant scatter. Part of the scatter arises from the \textsc{NSCs} for the reasons we were discussing above, namely, the complex star formation histories and mergers with other stellar systems. If we consider globular cluster systems, the simulations and observations appear very consistent and form a tight relation with efficiencies on the percent level at best. The \textsc{UCDs} are most uncertain, while the case of UCD2 (no dark matter, including expansion) fits best within the general trend, also the assumptions of UCD1 (dark matter, no expansion) fit into the general picture. We also emphasize that due to the unclear origin and the complex physics in these systems, a complete agreement could also be coincidental.

\section{Summary and conclusions}

In this paper, we investigate how the efficiency of black hole formation via collisions in stellar systems depends on the ratio of stellar over critical mass. The importance of the collision timescale in the formation of runaway massive objects has previously been suggested through an analysis of observational properties of both \textsc{NSCs} and massive black holes by \citet{Escala2021}, from the comparison of collision timescales with ages in galactic nuclei. The scenario was further tested with satisfactory results, using a set of idealized N-body simulations employing constant stellar masses by \citet{Vergara2023}, where a critical mass for collisions was introduced for the comparison of the collision time with the age of the system.

Here, we have generalized this concept to take into account the presence of an external potential and we employed the result from a large range of numerical simulations and observational data from the literature. This way we compared with the previous results and determined how the efficiency for the formation of massive objects depends on the ratio of initial stellar mass over the critical mass scale defined in Eq.~\ref{eq3_mass_crit_ext}. This included simulations with different types of radial profiles in their initial conditions \citep[e.g.,][]{Plummer1911, King1966, Miyamoto1975},  simulations employing different stellar IMFs \citep[e.g.,][]{Salpeter1955, Scalo1986, Kroupa2001} as well as simulations including stellar evolution \citep[][]{Hurley2000, Hurley2002} and simulations with one million stars \citep{Panamarev2019, Arca-Sedda2023a, Arca-Sedda2023b, Arca-Sedda2023c}. 

The comparison of these data is not completely trivial, for example, not all the output is always reported in the literature; also, while the initial conditions are usually well-described, the final stellar mass in the cluster and the mass loss due to escapers are not always given, but this still forms an important ingredient in our efficiency parameter, $\epsilon_{\mathrm{BH}}$. Nonetheless, even under the simplifying assumption that mass loss would be negligible, our analysis has shown a clear correlation between the efficiency required to form a massive object and the ratio of initial stellar mass over critical mass. Particularly, when that ratio is larger than 1, the typical efficiencies lie at least in the range of $20-50\%$; whereas for lower ratios, the most frequent outcome is within the percent range. We note that there is scatter in this relation and we also note some differences in the different sets of simulations, which are also unsurprising given the different assumptions involved. We further note that the obtained correlation improves if we derive a correction factor from the simulations of \citet{Reinoso2020, Vergara2021, Vergara2023}, as for those simulations, we know the fraction of stellar mass lost due to escapers. Assuming that the physics are similar in other simulations, the correlation becomes more clear and we find efficiencies of $20-60\%$ when the ratio of initial stellar over critical mass is greater than $1$.

Subsequently, we have analyzed observational data of dense stellar systems, including \textsc{GCs}, \textsc{NSCs}, and \textsc{UCDs}. 
The \textsc{GCs} are perhaps the clearest case as these are very old systems without any recent star formation and where the general evolution is very well understood. For these cases, the ratio of initial stellar over critical mass is in the range of $0.005-0.03$ and the efficiencies are very low $\sim 5\%$. The simulations of \citet{Portegies1999, Fujii2013, Mapelli2016} and \citet{Arca-Sedda2023a, Arca-Sedda2023b, Arca-Sedda2023c} show low $\epsilon_{\rm BH}$ and further substantiated by observation of \citet[e.g][]{Harris1996} and \citet[e.g][]{Lutzgendorf2011, Lutzgendorf2013a, Lutzgendorf2013, Lutzgendorf2013b}. This supports the scenario of \textsc{IMBHs} formation in \textsc{GCs}.

In the case of \textsc{NSCs}, we notice more scatter, as the systems are more complex including complicated star formation histories and typically also mergers with other stellar systems. They cover a wide range of initial over critical mass ratios of $0.01-100$ and, despite the presence of scatter, it is visible that very high efficiencies above $50\%$ are only reached for mass ratios above $0.3$. 

The \textsc{UCDs} are perhaps the most uncertain systems, as their nature and evolution are not fully clear. We have explored assumptions of these systems both assuming an evolution more like a \textsc{NSCs} and alternatively using an evolutionary path with a fixed radius but including an external potential due to dark matter. The first case is more favorable, while the second somewhat increases the scatter, although neither of these cases fundamentally change the picture. However, it is noticeable that the observational data do include more scatter than the simulations. This could be due to the higher complexity of those systems in general. Also, it is certainly not guaranteed that the observed systems formed just as a result of collisions, but it could be the case that additional gas physics was also involved, which could help to increase the efficiency with respect to forming a massive object \citep[e.g.,][]{Schleicher2022}. 

Overall, while it is clear that additional factors such as the complex history in real stellar systems and the possible presence of gas most likely have an influence on the results and may also be able to explain the scatter that we observe, we find efficiencies of $30-100\%$; these results are correlated with ratios of initial stellar over critical mass above 1. Observations reach higher efficiencies due to the secular evolutionary process of evaporation of the cluster. This result should not be particularly surprising, as it only means that the collision timescale should be short enough to play a relevant role within a given stellar system, but should be useful to determine the potential of a system to form very massive objects. It is worth stressing that high efficiencies can be reached for a wide range of systems. This opens up the potential to form quite massive black holes via the collision channel in massive \textsc{NSCs,} as long as their stellar mass is larger than their critical mass. 

In the future, it will be useful for this scenario to be tested further. For example, there are only a few N-body simulations with one million stars available, such as those of \citet{Panamarev2019, Arca-Sedda2023a, Arca-Sedda2023b, Arca-Sedda2023c}. More simulations and more compact systems will allow us to test this scenario further and under more realistic conditions. At the same time, the complexity of simulations could still be increased, by implementing ongoing star formation as expected within real \textsc{NSCs} and/or mergers with \textsc{GCs} as another possible growth channel. To understand the importance of this black hole formation mechanism, statistical predictions will be important in combining semi-analytic galaxy evolution models, as those of \citet[e.g.][]{Sassano2021, Trinca2022}, with black hole formation recipes employing the concept of the critical mass scale to obtain statistical predictions  at both high and low redshift. In the Local Universe, such predictions could be compared to the black hole mass function determined via the \href{https://webb.nasa.gov/}{\textit{James Webb} Space Telescope (JWST)} and the \href{https://elt.eso.org/}{Extremely Large Telescope (ELT)}, where new \textsc{IMBHs} may be discovered at the lower mass end and in general more data will become available for testing the scenario proposed here. At high redshift, this scenario can operate and its relevance for the formation of high redshift black holes needs to be further explored. 
 
While the potential importance of the collision-based channels for black hole formation has already been suggested by \citet{Rees1984}, previous estimates typically insinuate that they form black hole masses up to the order of $10^3$~M$_\odot$ \citep[e.g.,][]{Devecchi2009, Devecchi2010, Katz2015, Sakurai2017, Reinoso2018}. The results obtained here considerably strengthen and support the relevance of the collision-based channel, suggesting that it could efficiently operate in dense NSCs and the efficiency of SMBH formation in typical NSCs appears roughly consistent with this scenario, including SMBH masses on the order of $10^6$~M$_\odot$ and somewhat above. It is even further consistent with the very low efficiencies in GCs, while the results for UCDs are not fully conclusive, as they depend on certain assumptions being made concerning these objects. What we can conclude from the comparison with the large set of simulations analyzed in this work is that the scenario holds over a wide range of possible conditions, including variations of the radial profile of the clusters, the presence of an initial mass function, and the inclusion of additional physics such as stellar evolution. Future simulations  with high particle numbers and exploring additional physics, including improved treatments of collisions and stellar physics \citep[e.g.,][]{Kamlah2024}, and investigating the role of rotation and binaries will allow us to probe this scenario in even greater detail.

\begin{acknowledgements}
We thank the anonymous referee for his/her comments, which helped to improve the manuscript quality. MCV acknowledges funding through ANID (Doctorado acuerdo bilateral DAAD/62210038) and DAAD (funding program number 57600326). DRGS, AE and ML acknowledge the center for Astrophysics and Associated Technologies CATA (FB210003) and via ANID  Fondo 2022 QUIMAL 220002. DRGS and ML also acknowledge financial support from FONDECYT Regular grant $\#$1201280 and through the Alexander von Humboldt - Foundation, Bonn, Germany. BR acknowledges funding through ANID (CONICYT-PFCHA/Doctorado acuerdo bilateral DAAD/62180013), DAAD (funding program number 57451854), and the International Max Planck Research School for Astronomy and Cosmic Physics at the University of Heidelberg (IMPRS-HD). FFD, AWHK, and RS acknowledge financial support by German Science Foundation (DFG) project Sp 345/22-1 and Sp 345/24-1. Computations were performed on the HPC system Raven at the Max Planck Computing and Data Facility, and we also acknowledge the Gauss center for Supercomputing e.V. for computing time through the John von Neumann Institute for Computing (NIC) on the GCS Supercomputer JUWELS Booster at Jülich Supercomputing center (JSC). These resources made the presented work possible, by supporting its development. NH is a fellow of the International Max Planck Research School for Astronomy and Cosmic Physics at the University of Heidelberg (IMPRS-HD). NH acknowledges funding through \href{https://www.lacegal.com/}{LACEGAL}, a Latinamerican Chinese European Galaxy Formation Network. This project has received funding from the European Union’s HORIZON-MSCA-2021-SE-01 Research and Innovation programme under the Marie Sklodowska-Curie grant agreement number 101086388.
\end{acknowledgements}
%
\bibliographystyle{aa} 
\bibliography{ref} 
%

\begin{appendix} 
\section{Computation of the critical mass and black hole formation efficiency for simulations}\label{app_A}

\citet{Fujii2013} delved into the role of stellar collisions in ensemble star clusters and virialized solo-cluster models.  In this work, we focus on the latter. Those authors used a \citet{Salpeter1955} IMF with $M_{\rm min} = 1\mathrm{M_\odot}$ and $M_{\rm max} = 100\mathrm{M_\odot}$ and density profile of \citet{King1966}. To compute the critical mass and the black hole formation efficiency, we used the number of stars ($N$), the cluster mass ($M_{\mathrm{i}}$), the half-mass radius, the mean stellar mass ($\overline{m_*}=M_{\mathrm{i}}/N$), and, finally, the black hole mass after a simulation time of $5\,$Myr. In the study of \citet{Sakurai2017} they explore runaway collisions in the first star cluster, using a \citet{Salpeter1955} IMF varying the masses between $3-100\mathrm{M_\odot}$. The distribution of stars does not follow any profile mentioned before, since  the positions of the stars were obtained from previous hydrodynamics simulations. The parameters adopted to compute the $M_{\mathrm{crit}}$ and $\epsilon_{\mathrm{BH}}$ were the simulation time ($\tau =3\,$Myr), the initial cluster, the number of stars, and the black hole mass. 

\citet{Portegies1999} explored runaway collisions in young compact star clusters. They employed a \citet{Scalo1986} IMF, which varies the stellar mass between $0.1$ and $100 \, M_\odot$. We used $N$, $\overline{m_*}$, $M_{\mathrm{i}}=N*\overline{m_*}$, and $R_{\mathrm{h}}$ in addition to the values for the black hole mass and the simulation time (from Figure 2) to compute $M_{\text{crit}}$ and $\epsilon_{\text{BH}}$. \citet{Mapelli2016} explored the impact of metallicity in runaway collisions in young dense star clusters. The authors used the \citet{Kroupa2001} IMF with $M_{\rm min} = 0.1\mathrm{M_\odot}$ and $M_{\rm max} = 150\mathrm{M_\odot}$. The parameters used to calculate the critical mass and the black hole formation efficiency are $N$, $M_{\mathrm{i}}$, $\overline{m_*}=M_{\mathrm{i}}/N$, and $R_{\mathrm{h}}$. The stellar system was evolved for over $17\,$Myr, forming binary systems. We just consider the mass of the most massive one of the binary system. Both works used the \citet{King1966} profile.

\citet{Katz2015} investigated primordial star clusters at high redshift. They modeled both spherical \citep{Plummer1911} and non-spherical star distributions, but for our purposes, we selected the spherical model due to the availability of the virial radius ($R_v$). As per its definition, $R_v$ is not available for the non-spherical distribution, rendering it impossible to compute $M_{\text{crit}}$. The study used a \citep{Salpeter1955} IMF with $M_{\text{min}} = 1 \, M_\odot$ and $M_{\text{max}} = 100 \, M_\odot$. To calculate $M_{\text{crit}}$ and $\epsilon_{\text{BH}}$, we employed the following parameters $R_v$, $M_{\mathrm{i}}$, and $M_{\mathrm{BH}}$, although the value of $N$ is absent, they provided the IMF, which allowed us to calculate the mean stellar mass ($\overline{m_*}$) and subsequently determine the number of stars as $N = M_{\mathrm{i}} / \overline{m_*}$. The simulations run for $3.5 \, \text{Myr}$. \citet{Reinoso2018} explored collisions in primordial star clusters. They used equal-mass stars distributed according to a Plummer distribution \citep{Plummer1911}. To estimate $\epsilon_{\text{BH}}$ and $M_{\text{crit}}$, we used the initial cluster mass and the half-mass radius, the number of stars, the mean stellar mass $\overline{m_*} = M_{\mathrm{i}} / N$, and stellar radii $r_* = 20, 50, 100, 200, 500, 1000, 5000 \, R_\odot$. The larger stellar radii were assumed based on a high accretion rate of $0.1 \, M_\odot \, \text{yr}^{-1}$ for the protostars \citep{Hosokawa2012, Hosokawa2013, Schleicher2013, Haemmerle2018}. Their simulations spanned $2 \, \text{Myr}$ and resulted in the formation of very massive stars. These simulations do not include stellar evolution, however, if the simulation time is sufficient to form a black hole, then they consider the masses of these very massive stars as the black hole mass. \citet{Reinoso2020} investigated runaway collisions in dense star clusters, considering both the presence and absence of external potential effects. They adopted a star distribution from \citet{Plummer1911} with equal-mass stars. To compute the critical mass, we use Eq.~\ref{eq3_mass_crit} for systems without external potential and Eq.~\ref{eq3_mass_crit_ext} for systems with external potential, the black hole formation efficiency was computed with Eq.~\ref{eff_bh}. To compute these quantities, we use $R_v$ (the external potential also had the same $R_v$); $M_{\mathrm{i}}$; $N$; the mean stellar mass ($\overline{m_*} = M_{\mathrm{i}} / N$); and $r_*$. The above-mentioned simulations extended over $10 \, \text{Myr}$ and resulted in the formation of very massive stars. They used the same assumption as in \citet{Reinoso2018} for the larger stellar radii, with the very massive stars ending up as a black hole. Neither \citet{Katz2015} nor \citet{Reinoso2020} provided a value for $R_{\mathrm{h}}$, but they did provide the value for $R_v$. To approximate $R_{\mathrm{h}}$, we used the relation typical for a Plummer profile, a virial radius of approximately $R_v \approx 1.7a$, and a half-mass radius of approximately $R_{\mathrm{h}} \approx 1.3a$, where $a$ is the Plummer radius, leading the following estimation of $R_{\mathrm{h}} \approx 0.8R_v$. \citet{Vergara2021} investigated runaway collisions in both flat and rotating clusters using a Miyamoto-Nagai distribution \citep{Miyamoto1975} with equal-mass stars. To compute the $M_{\mathrm{crit}}$ and $\epsilon_{\mathrm{BH}}$, we used the initial cluster mass and a half-mass radius of the radial plane,  numbers of stars, and  stellar mass as $m_* = M_{\mathrm{i}} / N$, with a range of stellar radii, $r_* = 50, 100, 500, 1000 \, R_\odot$ (with the stellar radius used as a free parameter). The simulation time was $2 \, \text{Myr}$ and resulted in the formation of very massive stars. They also assumed these very massive stars will end as a black hole.  

\citet{Panamarev2019} simulated the galactic center of the Milky Way, using one million stars, including a \citet{Kroupa2001} IMF with masses ranging from $0.08$ to $100 \, M_\odot$. To assess the efficiency of black hole formation and the critical mass, we used an initial mass,  half-mass radius,  time ($\tau =5.5 \, \text{Gyr}$), and the black hole mass  taken from their Figure 7. In a different study, \citet{Vergara2023} investigated runaway collisions in \textsc{NSCs}, employing models with equal-mass stars distributed according to a Plummer distribution \citep{Plummer1911}. To determine the critical mass and black hole formation efficiency, we considered the initial cluster mass, virial radii,  stellar mass and radius, and the mass of very massive stars formed after $\tau =10 \, \text{Myr}$. In \citet{Reinoso2018, Reinoso2020, Vergara2021}, the authors assumed that very massive stars evolve enough time to form a black hole. We computed the half-mass radius using the same assumptions as \citet{Reinoso2020} and \citet{Katz2015}. Besides \citet{Arca-Sedda2023a, Arca-Sedda2023b, Arca-Sedda2023c} simulated star cluster using $(0.12,0.3,0.6,1)\times 10^6$ stars, including a \citet{Kroupa2001} IMF ranging from $0.08$ to $150 \, M_\odot$. These simulations also include primordial binaries up to $33\%$ of the total number of stars. The simulations last around a few gigayears. We took the maximum black hole mass formed in the stellar system (see their Table 1).

\citet{Rizzuto2023} explored tidal disruption and capture runaway collisions in dense clusters, including post-Newtonian effects. The use of a \citet{Kroupa2001} IMF, with a range between $0.08$ to $2 \, M_\odot$, the low-mass stars were chosen to avoid stellar evolution. They also include an initial black hole within the cluster with initial masses of $50$, $300,$ and, $2000 \, M_\odot$. Almost all the simulations run for times around $40$\, Myr. To compute the efficiency of black hole formation and the critical mass, we used an initial mass, the half-mass radius, the simulation time, and the black hole mass.

\section{Cluster mass loss}\label{ClusterMassLoss}

The relation between the initial mass, $M_{\mathrm{i}}$, and final mass, $M_{\mathrm{f}}$, of stellar systems is not always well known. Simulations studying the formation of massive objects via collisions usually report the initial stellar mass of the system as well as the mass of the massive object that forms; sometimes, also additional information on the final stellar mass and the escapers is being reported. Observations, on the other hand, only report on the current state of a stellar system. For the analysis in this paper, the relation between final and initial stellar mass can, however, be somewhat relevant and, therefore, this aspect is examined here in some detail.

As a first and very simple approximation, we could assume that the total mass in the system is conserved, namely, that there are no escapers and the mass exchange happens only between the stellar mass and the central massive object as a result of collisions. In this case, we have:

\begin{equation}\label{simple}
    M_{\mathrm{i}}=M_{\mathrm{f}}+M_{\mathrm{BH}}.
\end{equation}

\begin{figure}
    \centering
    \includegraphics[width=1\linewidth]{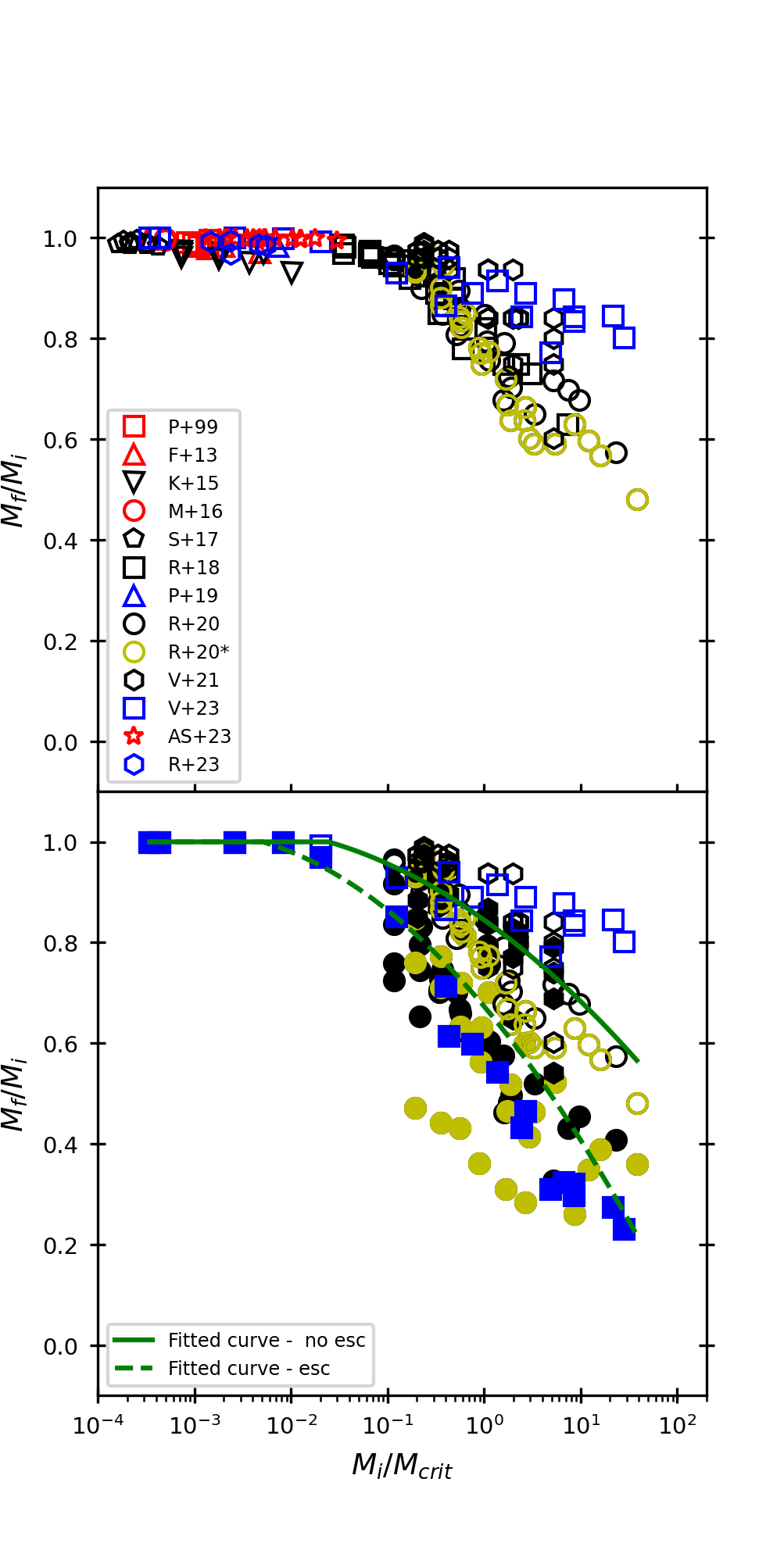}
    \caption{$M_{\mathrm{f}}/M_{\mathrm{i}}$ as a function of $M_{\mathrm{i}}/M_{\mathrm{crit}}$, assuming that $M_{\mathrm{i}}=M_{\mathrm{f}}+M_{\mathrm{BH}}$ for all simulations considered from Table~1 (top).  $M_{\mathrm{f}}/M_{\mathrm{i}}$ as a function of $M_{\mathrm{i}}/M_{\mathrm{crit}}$ (bottom) using the simulations of \citet{Reinoso2020, Vergara2021, Vergara2023}. We show both the result as it would appear considering the approximation used in Eq.~\ref{simple} via the solid line and the result when we employ the initial and final stellar masses obtained from the simulation. In the second case the plot is qualitatively similar but with a steeper decrease due to the escapers. The figure also indicates our polynomial fits to the obtained relations.}
    \label{fig_MfMi_all_sims}
\end{figure}

In the case of the simulations, $M_{\mathrm{i}}$ and $M_{\mathrm{BH}}$ are known. The critical mass, $M_{\mathrm{crit}}$, can be calculated from the initial conditions and $M_{\mathrm{f}}$ can be estimated from Eq.~\ref{simple}. Employing this approximation, this allows us to plot $M_{\mathrm{f}}/M_{\mathrm{i}}$ as a function of $M_{\mathrm{i}}/M_{\mathrm{crit}}$ for the simulations, which we provide in the top panel of Fig.~\ref{fig_MfMi_all_sims}. Under this assumption, we see clearly that for small ratios of $M_{\mathrm{i}}/M_{\mathrm{crit}}\lesssim0.1$, $M_{\mathrm{f}}/M_{\mathrm{i}}\sim1$, while it decreases for higher values of $M_{\mathrm{i}}/M_{\mathrm{crit}}$. This behavior should indeed be expected if the mass of the most massive object increases with $M_{\mathrm{i}}/M_{\mathrm{crit}}$.

We can test the approximation employed in Eq.~\ref{simple}, at least for some simulations where both the initial and final stellar masses are available. For this purpose, we can employ the numerical simulations by \citet{Vergara2021, Vergara2023} introduced above, for which we  plot the relation both using the correct values from the simulation, and comparing with the result when instead using the approximation from Eq.~\ref{simple}. In the bottom panel of Fig.~\ref{fig_MfMi_all_sims}, we show that the results in both cases are qualitatively similar, however, when using the real data, the ratio $M_{\mathrm{f}}/M_{\mathrm{i}}$ decreases somewhat more rapidly with increasing $M_{\mathrm{i}}/M_{\mathrm{crit}}$ due to the escapers. This result also intuitively makes sense, as a higher ratio of $M_{\mathrm{i}}/M_{\mathrm{crit}}$ implies a higher probability that interactions within the cluster have occurred, which can also lead to the ejection of stars from the system. From the data obtained here, we derive a polynomial fit to determine $M_{\mathrm{f}}/M_{\mathrm{i}}$ as a function of $M_{\mathrm{i}}/M_{\mathrm{crit}}$. Our fit in the first case when employing Eq.~\ref{simple} is given as
\begin{equation}\label{fit_eq_no_esc}
    \left(\frac{M_{\mathrm{f}}}{M_{\mathrm{i}}}\right)_{\rm no esc} = a\log{\left(\frac{M_{\mathrm{i}}}{M_{\mathrm{crit}}}\right)^2} + b\log{\left(\frac{M_{\mathrm{i}}}{M_{\mathrm{crit}}}\right)} + c,
\end{equation}
while when using the real data including escapers, we obtain
\begin{equation}\label{fit_eq_esc}
    \left(\frac{M_{\mathrm{f}}}{M_{\mathrm{i}}}\right)_{\rm esc} = d\log{\left(\frac{M_{\mathrm{i}}}{M_{\mathrm{crit}}}\right)^2} + e\log{\left(\frac{M_{\mathrm{i}}}{M_{\mathrm{crit}}}\right)} + f.
\end{equation}

\begin{figure}
    \centering
    \includegraphics[width=1\linewidth]{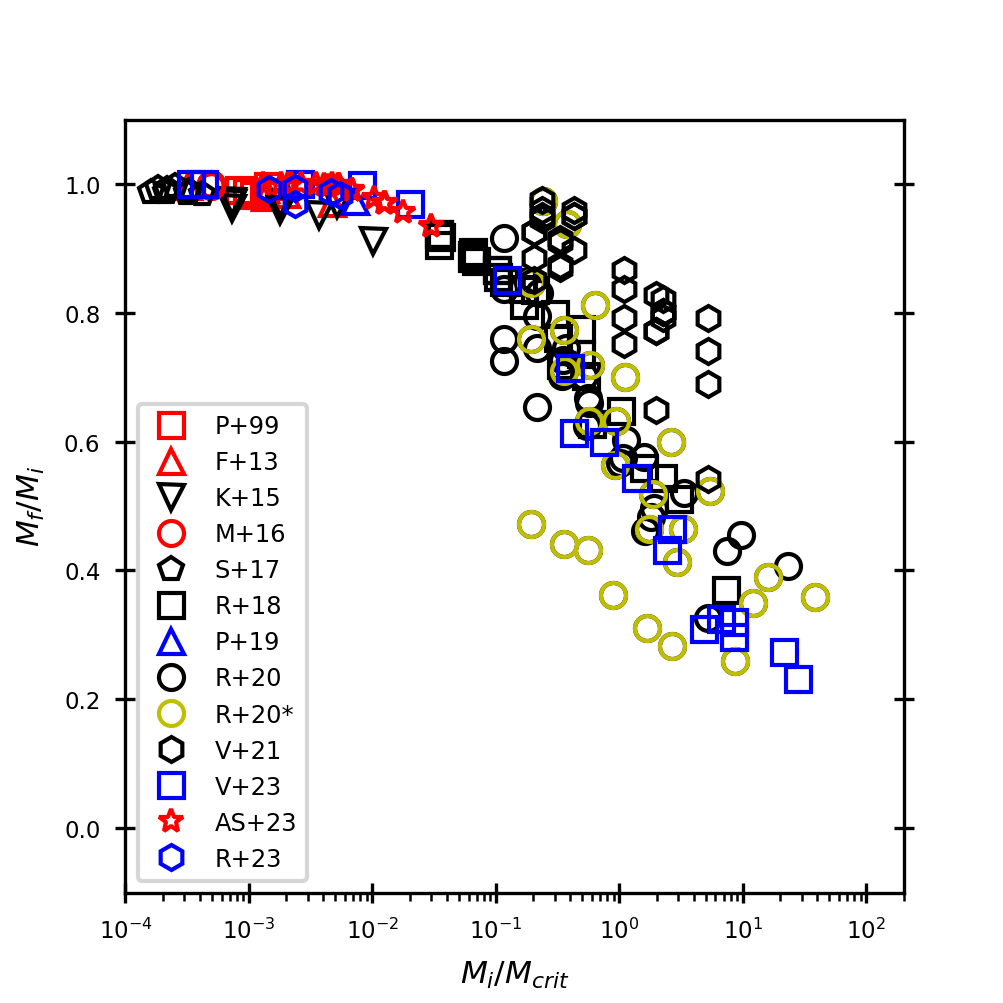}
    \caption{$M_{\mathrm{f}}/M_{\mathrm{i}}$ as a function of $M_{\mathrm{i}}/M_{\mathrm{crit}}$ in the simulations. Here $M_{\mathrm{f}}/M_{\mathrm{i}}$ is calculated considering the stellar mass going into the most massive object and applying the correction factor $\alpha$ from Eq.~\ref{alpha} to estimate the mass loss due to escapers.}
    \label{fig_MfMi_fit_all_sims}
\end{figure}

From the ratio between these two expressions, we can further derive a scale factor of 

\begin{equation}
\alpha=\left( M_{\mathrm{f}}/M_{\mathrm{i}} \right)_{\rm no esc}-\left( M_{\mathrm{f}}/M_{\mathrm{i}} \right)_{\rm esc}\label{alpha}
,\end{equation}

which can be employed as an approximate correction to estimate the effect of escapers when the information is not given. This procedure allows us to precisely take into account the mass exchange between mass in stars and the mass of the most massive object according to the simulation results, while the mass loss due to escapers is only estimated. The result of this estimate is given in Fig.~\ref{fig_MfMi_fit_all_sims}. We note that it shows a similar behavior as found in Fig.~\ref{fig_MfMi_all_sims}, which, in principle, is natural as the basic trend is seen even when  the mass loss due to escapers is not considered; this trend  is then  further enhanced when that possibility is taken into account. 

\end{appendix}

\end{document}